 \documentclass{article}

\usepackage{epsfig} 
\usepackage{amsmath}
\usepackage{amsfonts}
\usepackage{graphicx}

 \usepackage{amssymb}

\date{\today} 

\begin{document}

\title{
Generalized dyons and magnetic dipoles:\\
the issue of angular momentum
}

\author{
{\large Francisco Navarro-L\'erida,}$^{\ddagger}$
{\large Eugen Radu}$^{\diamond}$
and {\large D. H. Tchrakian}$^{\star\dagger}$ 
\\
\\ 
$^{\ddagger}${\small Dept.de F\'isica At\'omica, Molecular y Nuclear, Ciencias F\'isicas,}\\
{\small Universidad Complutense de Madrid, E-28040 Madrid, Spain}\\  
$^{\diamond}${\small Departamento de F\'\i sica da Universidade de Aveiro and I3N, 
   Campus de Santiago, 3810-183 Aveiro, Portugal}\\ 
$^{\star}${\small School of Theoretical Physics, Dublin Institute for Advanced Studies, 10 Burlington
Road, Dublin 4, Ireland }\\
{$^{\dagger}$\small Department of Computer Science, NUI Maynooth, Maynooth, Ireland}}

\date{}
\newcommand{\dd}{\mbox{d}}
\newcommand{\tr}{\mbox{tr}}
\newcommand{\la}{\lambda}
\newcommand{\ka}{\kappa}
\newcommand{\f}{\phi}
\newcommand{\vf}{\varphi}
\newcommand{\F}{\Phi}
\newcommand{\al}{\alpha}
\newcommand{\ga}{\gamma}
\newcommand{\de}{\delta}
\newcommand{\si}{\sigma}
\newcommand{\Si}{\Sigma}
\newcommand{\bnabla}{\mbox{\boldmath $\nabla$}}
\newcommand{\bomega}{\mbox{\boldmath $\omega$}}
\newcommand{\bOmega}{\mbox{\boldmath $\Omega$}}
\newcommand{\bsi}{\mbox{\boldmath $\sigma$}}
\newcommand{\bchi}{\mbox{\boldmath $\chi$}}
\newcommand{\bal}{\mbox{\boldmath $\alpha$}}
\newcommand{\bpsi}{\mbox{\boldmath $\psi$}}
\newcommand{\brho}{\mbox{\boldmath $\varrho$}}
\newcommand{\beps}{\mbox{\boldmath $\varepsilon$}}
\newcommand{\bxi}{\mbox{\boldmath $\xi$}}
\newcommand{\bbeta}{\mbox{\boldmath $\beta$}}
\newcommand{\ee}{\end{equation}}
\newcommand{\eea}{\end{eqnarray}}
\newcommand{\be}{\begin{equation}}
\newcommand{\bea}{\begin{eqnarray}}

\newcommand{\ii}{\mbox{i}}
\newcommand{\e}{\mbox{e}}
\newcommand{\pa}{\partial}
\newcommand{\Om}{\Omega}
\newcommand{\vep}{\varepsilon}
\newcommand{\bfph}{{\bf \phi}}
\newcommand{\lm}{\lambda}
\def\theequation{\arabic{equation}}
\renewcommand{\thefootnote}{\fnsymbol{footnote}}
\newcommand{\re}[1]{(\ref{#1})}
\newcommand{\R}{{\rm I \hspace{-0.52ex} R}}
\newcommand{\N}{{\sf N\hspace*{-1.0ex}\rule{0.15ex}%
{1.3ex}\hspace*{1.0ex}}}
\newcommand{\Q}{{\sf Q\hspace*{-1.1ex}\rule{0.15ex}%
{1.5ex}\hspace*{1.1ex}}}
\newcommand{\C}{{\sf C\hspace*{-0.9ex}\rule{0.15ex}%
{1.3ex}\hspace*{0.9ex}}}
\newcommand{\eins}{1\hspace{-0.56ex}{\rm I}}
\renewcommand{\thefootnote}{\arabic{footnote}}

\maketitle


\bigskip

\begin{abstract}
It is known that a non-Abelian magnetic monopole  cannot rotate globally
(although it may possess a nonzero angular momentum density).
At the same time, the total angular momentum
of a magnetic dipole equals the electric charge.
In this work we question the generality of these results by considering
a number of generalizations of the Georgi-Glashow model.
We study two different types of finite energy, regular
configurations: solutions with net magnetic charge and monopole-antimonopole
pairs  with zero net magnetic charge. 
These configurations are endowed with an electric charge
and carry also a nonvanishing angular momentum density.
However, we argue that the qualitative results found in the  Georgi-Glashow model 
are generic and thus a magnetic monopole cannot spin 
as long as the matter fields feature the usual ``monopole''  asymptotic behaviour
  independently of the dynamics of the model.
A study of the properties of the dyons and magnetic dipoles
in some generalizations of the Georgi-Glashow model supplemented
with higher order  Skyrme-like terms in the gauge curvature and Higgs fields
is given quantitatively.
 
\end{abstract}
\medskip
\medskip

\section{Introduction and motivation}

The existence of soliton solutions is an interesting feature of
some nonlinear field theory models.
Solitons behave like particles,
the fields being localized in smooth concentrations of energy density.
Moreover, some   static  solitons
are topologically stable
(see \cite{book} for a review of these aspects).
The systematic study of solitons
can be traced back at least to the work of Skyrme~\cite{Skyrme}.
The Skyrme model contains scalar fields (subject to a constraint) only, 
being proposed as an effective theory for nucleons.  In $3+1$ dimensional Minkowski spacetime (the case of interest in this work),
solitons exist also in models featuring non-Abelian gauge fields\footnote{It it worth noting that
no regular particle-like solutions of the pure Yang-Mills (YM)
equations exist in Minkowski spacetime  \cite{Deser:1976wq}. 
Physically this can be understood
as a consequence of the repulsive nature of the YM vector fields. 
(When scalar fields are added, 
the existence of solitons become possible due to the balance of
the YM repulsive force and the attractive force of the scalars.)
However, the no-go results in  \cite{Deser:1976wq}
are circumvented when including gravity effects,
as shown by the Bartnik and McKinnon (BK) family of solutions
with YM matter fields only 
\cite{Bartnik:1988am}.}.
The most prominent such configurations are the 't~Hooft-Polyakov magnetic monopoles in the Georgi-Glashow (GG)
model~\cite{'tHooft:1974qc}, \cite{Polyakov:1974ek}.   Both Skyrmions and monopoles~\cite{'tHooft:1974qc,Polyakov:1974ek}
are static topologically stable solitons. There exist also a tower of (excited) monopoles~\cite{Tchrakian:2010ar}, generalizing the monopoles of the GG model,
each arising from the dimensional descent of the $p$-th member of the Yang-Mills hierarchy~\cite{Tch} on $\R^3\times S^{4p-3}$ to $\R^3$. 

 Contrasting with these, are the sphalerons in the electroweak sector of the standard model \cite{Klinkhamer:1984di}, which have finite
energy but are not topologically stable. In the present work, this distinction between topologically stable and sphaleron-like solutions
will play a central role. 

The 't Hoof-Polyakov monopoles 
and their Julia-Zee dyonic generalizations \cite{JZ}
are static and spherically symmetric.
Then it is natural to wonder whether they posssess axially symmetric
generalizations with nonzero angular momentum.
This question has recently been addressed in the literature,
finding a rather unexpected answer. First, it turns out 
\cite{VanderBij:2001nm},
\cite{Kleihaus:2005fs} 
that the total angular momentum of the solitons endowed with a 
net magnetic (topological) charge vanishes (despite the fact that their angular
momentum density can be nonzero). 
Second, the angular momentum of a spinning magnetic dipole (or, more generally, of a configuration
with a vanishing net magnetic charge) is nonzero \cite{Paturyan:2004ps}
being proportional with the total electric charge.

The pivotal mechanism leading to this conclusion is the fact that the angular momentum
density becomes a total divergence by virtue of the electric component of the
Euler-Lagrange equations, and, the resulting surface integral then yields vanishing
global angular momentum when magnetic monopole boundary values are applied.

  The central physical question which we propose to address in 
this work concerns the generality of these results, namely that solitons with nonvanishing magnetic
monopole charge do not spin, irrespective of the specific dynamical model in question.
In practice, the  boundary values mentioned above, result from the presence of standard, {\bf $i.e.$ quadratic} 'kinetic' terms
of the Higgs and Yang-Mills fields in the Lagrangian.
However, 
modifications of standard field theoretic models can result in new non-perturbative effects. Examples of
this in $3+1$ dimensions are, for example, $i)$ gravity coupled to nonlinear electrodynamics \cite{AyonBeato:1998ub}, which
results in regular black hole,  $ii)$ the existence of glueballs in a non-Abelian Born-Infeld theory \cite{Gal'tsov:1999vn} and
$iii)$ the stabilising of black holes in Einstein-Yang-Mills theory augmented with higher order Yang-Mills (YM) curvature terms \cite{Radu:2011ip}.
Moreover, in higher dimensions, the systematic introduction of higher order YM curvatures and their dimensional descendants
results in the construction of instantons and monopoles in all (possible) dimensions~~\cite{Tchrakian:2010ar}.

 Seeing that more general models than the standard ones with quadratic kinetic terms only may result 
in qualitatively different solutions, one cannot $a\ priori$ exclude the
possibility that  
a more general choice of the Yang-Mills-Higgs
(YMH) Lagrangian might invalidate the results in 
\cite{VanderBij:2001nm}, 
\cite{Kleihaus:2005fs} 
on the general connection between the angular momentum and the
electric and magnetic charges.  
Moreover, by considering generic extensions of YMH systems, we hope to  shed some light on the problem of spinning
solitons with non-Abelian fields in general. 

In what follows, we shall address this question
by considering several different generalizations of the 
GG model and computing the corresponding 
angular momentum for a given choice  of boundary conditions 
at infinity, reflecting the presence or not of a magnetic charge.
Quantitative results are also shown for a specific choice of the YMH Lagrangian.
Within the terminology in this work,
the configurations with a model with net magnetic and electric charges are called {\it generalized dyons}.
We shall also consider composite configurations with an electric charge only, corresponding
to   {\it generalized dipoles}. 

This paper is structured as follows.
A discussion of the general aspects of the problem of spinning
solutions with YMH fields
is given in Section 2,
where we exhibit the general framework of the problem.
A new result there is a proof that the total angular momentum of a generic YMH system,
evaluated by employing the canonical energy momentum tensor,
can be expressed as a surface integral at infinity.
We continue with Section 3, where we discuss
the basic properties of several generalizations of  the GG model.
Most of the work in this paper is done for a  model 
where the usual the GG Lagrangian is 
supplemented with higher order curvature terms of the gauge field and 
 Skyrme-like terms of the gauged Higgs.
 Our choice of the supplementary part of the 
 Lagrangian has a natural justification,
 since it can be viewed as the second term ($p=2$) in a hierarchy of models
descended from $4p$ dimensional YM systems \cite{Tch}, the first one of which, $p = 1$, is the
usual GG model in the Prasad-Sommerfield (PS) limit.
Apart from that, we shall consider also a YMH model with the quadratic 
YM Lagrangian replaced by a non-Abelian Born-Infeld term,
and a  YMH  model recently introduced in \cite{Navarro-Lerida:2013pua},
featuring an extra Chern-Simons--like term providing a supplementary
interaction between the gauge and Higgs sectors of the theory.
The main result of this work is contained in Section 4, where we argue that
the general connection between the angular momentum and the
electric and magnetic charges 
found in
\cite{VanderBij:2001nm},
\cite{Kleihaus:2005fs} 
for the GG model,
still holds as long as asympototically  the  Higgs field approaches
a constant (non-zero) value and the gauge derivative of the Higgs field
vanishes.
In particular,
a magnetic  monopole 
does not possess generalizations with a nonzero total angular momentum.  
We continue in  Section 5 with a quantitative 
study of the generalized dyons
and  generalized dipoles
in some limits of the $(p=1)+(p=2)$ YMH model.
By solving numerically the corresponding field equations, we study how these
known solutions of the GG model are affected by the presence of higher derivative YMH terms
in the Lagrangian. 
The last Section of this work is devoted to a summary and 
a discussion of our results.
Several possible extensions of the results in Section {\bf 4}
are also mentioned there. In an Appendix, we present a peculiar electrically neutral solution of a system consisting
of the usual $F^2$ YM term, plus the pure $p=2$ YMH system, which might signal the circumvention of
the no-angular momentum conjecture. However, it is shown there that
 the ``dyon'' of this system must exhibit ``magnetic monopole'' boundary
values and hence the ban on angular momentum for topologically stable YMH solitons cannot be circumvented in this way,
supporting our explanation in Section {\bf 6}.


\section{The formalism}

\subsection{The conventions and notations}

In this work we shall ignore the backreaction of the matter 
fields on the geometry and consider a fixed four dimensional
Minkowski spacetime background\footnote{See, however, the remarks  in the last Section,
on the gravity
effects.},
with a line element 
$ds^2=dt^2-(dx^2+dy^2+dz^2)$, where $t$ is the time coordinate and $x,y,z$
are the usual cartesian coordinates.
The same line element expressed in spherical coordinates reads
$ds^2=dt^2-dr^2-r^2(d\theta^2+\sin^2\theta d\varphi^2)$, where
$0\leq r<\infty$ is the radial coordinate and $\theta,\varphi$ are
the spherical coordinates on $S^2$, with the usual range.
For completness, we give also corresponding expression in cylindrical coordinates, 
$ds^2=dt^2-d\rho^2-\rho^2d\varphi^2-dz^2$, where $0\leq \rho<\infty$.
Note that in all relations, the greek indices (like $\mu,\nu$) are running from $0$ to $3$ (with $x^0=t)$.

The gauge potential $A_{\mu}$ and Higgs field $\Phi$ are denoted as
\be
\label{0}
A_\mu =-\frac{i}{2}A_\mu^a \tau_a \: , \qquad \Phi = -\frac{i}{2}\Phi^a \tau_a,
\ee
with $\tau_a$ the Pauli matrices ($a=1,2,3$).
 The resulting antihermitian curvature and
covariant
derivative are then
\be
\label{c}
F_{\mu\nu} =\pa_\mu A_\nu -\pa_\nu A_\mu +[A_\mu ,A_\nu]\: ,
~~
D_\mu \Phi =\pa_\mu \Phi +[A_\mu ,\Phi].
\ee
Also, for the purposes of this work we found convenient to introduce the notation
\bea
\nonumber
\label{cal0}
{\cal F}_{\mu\nu\rho\si}&=&\{F_{\mu[\nu},F_{\rho\si]}\}=F_{\mu\nu\rho\si},
\\
\nonumber
{\cal F}_{\mu\nu\rho}&=&\{F_{[\mu\nu},D_{\rho]}\F\},
\label{cal1}
\\
{\cal F}_{\mu\nu}&=&\{S,F_{\mu\nu}\}+[D_{\mu}\F,D_{\nu}\F],
\label{cal2}
\\
\nonumber
{\cal F}_{\mu}&=&\{S,D_{\mu}\F\},
\label{cal3}
\\
\nonumber
{\cal F}&=&S^2\label{cal4},~~~{\rm with}~~S\stackrel{\rm def}=-(\eta^2\eins+\F^2).
\eea
where
we have used $[ijk]$ to denote cyclic symmetry in the indices $i, j, k$ and 
$\{A,B\}$ denotes an anticommutator.
The generalizations the GG Lagrangian 
we shall consider in what follows
are built in term of the terms (\ref{cal2}) above.

\subsection{The issue of axial symmetry}
The physical configurations we are interested in
have  no time dependence and are axially symmetric 
($i.e.$ they remain invariant under a rotation around the $z-$axis).
This implies the existence of two Killing vectors of the problem 
$\partial/\partial t$
and $\partial/\partial \varphi$.
For an Abelian gauge field this implies directly that the four potential
has no dependence on $t$ and $\varphi$.
However, the issue of interplay between spacetime and gauge symmetries
is rather subtle in the presence of non-Abelian matter fields\footnote{A nice discussion of the 
relationship between
conservation laws, spacetime symmetries and gauge symmetries can be found in Ref. \cite{Volkov:2003ew}.}.
In this case,
the symmetry of the gauge field under a spacetime symmetry (as characterized by a given Killing vector)
 means that the action of an isometry
can be compensated by a suitable gauge transformation \cite{Heusler:1996ft,Forgacs:1980zs}.
  For the Killing vector $\partial/\partial t$, 
  the natural choice is to choose a gauge with
 $\partial A/\partial t=\partial \Phi/\partial t=0$.
However, a rotation around the $z-$axis
can be compensated by a gauge rotation.
$
\partial_\varphi A_\mu=D_\mu\psi,
$
(with $\psi$ an element of the algebra),
and therefore
\begin{eqnarray} 
\label{relations}
F_{\mu \varphi} =  D_{\mu}W,
~~
D_{\varphi}\Phi= [W,\Phi],
\end{eqnarray}
where
\be
\label{W}
W=A_{\varphi}+\psi\,.
\ee
These relations (which are independent of the choice of a specific YMH Lagrangian),
allow us in what follows to express the angular momentum density as a flux integral at infinity.

\subsection{The energy-momentum tensor and  general relations}
The Lagrangian ${\cal L}$ of the model (which is not specified at this stage)
is a function of the matter fields $\Psi=(A_{\mu},\Phi)$.
However, the requirement that the equations of motion are of second order
plus the gauge covariance
implies that ${\cal L}$ depends only on $F_{\mu \nu}$, $D_{\mu}\Phi$
and $\Phi$.

We start by defining the generalized momentum densities 
\begin{eqnarray}
\label{def1}
\Pi^{\mu\nu}=\frac{\partial {\cal L} }{\partial  A_{\nu,\mu}},~~
\Pi^\mu=\frac{\partial {\cal L}}{\partial  \Phi_{,\mu}},
\end{eqnarray} 
such that the Euler-Lagrange equations $\delta_\Psi {\cal L}=0 $ can be written as\footnote{Note that these relations are
given in a Cartesian coordinate system.}
\begin{eqnarray}
\label{YMH-eqs}
D_{\mu}\Pi^{\mu \nu}-[\phi,\Pi^\nu]=0,~~
\Pi^{\mu}_{,\mu}=\frac{\partial {\cal L}}{\partial \Phi}.
\end{eqnarray}
As usual, the invariance of the action under  four-translations,
$x^\mu \to x^\mu +a^\mu$ yields the 
 canonical energy-momentum tensor \cite{LL}
\begin{eqnarray}
\label{canonical-Tik}
2T_\alpha^\beta= 
{\rm Tr} 
\bigg(
A_{\mu,\alpha} \frac{\partial {\cal L}}{\partial A_{\mu,\beta}}
+\Phi_{,\alpha}\frac{\partial {\cal L}}{\partial \Phi_{,\beta}}
\bigg )
-\delta_\alpha^\beta {\cal L}
=
{\rm Tr} 
\bigg(
A_{\mu,\alpha} \Pi^{\beta \mu}+
\Phi_{,\alpha }\Pi^{\beta }
\bigg )
-\delta_\alpha^\beta {\cal L}  
,
\end{eqnarray}
which is conserved, $T_{\mu,\nu}^\nu=0$.
Following the usual prescription,
this expression can be made gauge invariant by adding to it a total divergence \cite{Belinfante}, 
such that
\begin{eqnarray}
\label{canonical-Tik-1}
2T_\alpha^\beta = 
{\rm Tr} 
\bigg(
F_{\alpha\mu } \Pi^{ \beta \mu}
+D_{\alpha}\Phi\Pi^\beta
\bigg)
-\delta_\alpha^\beta {\cal L}
.
\end{eqnarray}

At this point is is worth recalling that the canonical energy momentum tensor $T_{\mu \nu}$
suffers from a number of well-known problems;
for example it is not explicitly symmetric in the indices $\mu,\nu$ 
(this holds also for (\ref{canonical-Tik-1})).
As usual, an energy-momentum tensor which is directly symmetric and gauge invariant 
is found by introducing the space–time metric $g_{\mu\nu}$ into
the action and assuming it to be arbitrary;
then the energy-momentum tensor is obtained
by differentiating the density of the action with respect to the metric:
\begin{eqnarray}
\label{metric-Tik}
 T_{\alpha \beta}
=  \frac{2}{\sqrt{-g}}
\frac{\delta \sqrt{-g} {\cal L}}{\delta g^{\alpha \beta}}.
\end{eqnarray}
One can easily show that  the energy-momentum tensor and the canonical one 
are identical for the case of a GG model. 
However, 
we have verified that the expression of $T_{\alpha \beta}$
obtained via the definition  
(\ref{canonical-Tik-1})
and via (\ref{metric-Tik})
also  coincide 
for the specific models 
(\ref{gen-L}), 
(\ref{LBIH})
and 
(\ref{L-YMHCS})
 bellow.

 The advantage of using the definition (\ref{canonical-Tik-1})
 is that it leads to an expression of the angular momentum density, $T_\varphi^t$,
 as a total divergence, independent on the choice of ${\cal L}$.
In proving that, we make use of the general relations (\ref{relations}) 
together with the generalized YM equations (\ref{YMH-eqs}).
After replacing in 
one finds the following general expression\footnote{Here we 
change to a spherical (or cylindrical) coordinate system, such that $\mu=r,\theta$ 
(or  $\mu=\rho,z$, respectively), $g$ being the determinant of the metric tensor.} of  $T_\varphi^t$
\begin{eqnarray}
\label{T34}
2T_\varphi^t
= 
~{\rm Tr} \bigg(
\frac{1}{\sqrt{-g}}
\frac{\partial}{\partial  x^\mu }
(
\sqrt{-g}
 W \Pi^{\mu t} 
 )
 \bigg).
\end{eqnarray}
As a result, the total angular momentum
can be expressed as a flux integral\footnote{Note  that  the eq. (\ref{J}) 
holds also for higher gauge groups, since it relies on the general relation (\ref{relations}).} 
\begin{eqnarray}
\label{J}
J= \int d^3 x \sqrt{-g} T_\varphi^t
=
\frac{1}{2}\oint_\infty d\Sigma_k 
~{\rm Tr}(
\sqrt{-g}
 W \Pi^{k t} 
 ).
\end{eqnarray}
Therefore we conclude that in a general YMH theory
only the large$-r$
asymptotic structure of the fields is relevant for the issue of 
angular momentum of regular configurations.
(Note that this relation has been proven in 
\cite{VanderBij:2001nm} 
for a GG model.) 

For the purposes of this work,
it is also useful to define the {\it 'electric'}
part of the energy density as
\begin{eqnarray}
\label{electric1}
2{\cal E}={\rm Tr} 
\bigg(
F_{\alpha t } \Pi^{ \alpha t}
+D_{t}\Phi\Pi^t
\bigg)
.
\end{eqnarray}
Similar to the angular momentum density, one can show that (\ref{electric1}) can also be written
as total divergence
\begin{eqnarray}
\label{electric2}
2{\cal E}=
~{\rm Tr} \bigg(
\frac{1}{\sqrt{-g}}
\frac{\partial}{\partial  x^\mu }
(
\sqrt{-g}
A_t \Pi^{\mu t} 
 )
 \bigg).
 \end{eqnarray}
Then the total {\it 'electric' mass} can also be expressed
as a surface integral
\begin{eqnarray}
\label{E}
E_e=\int d^3 x \sqrt{-g} {\cal E}
=
\frac{1}{2}\oint_\infty d\Sigma_k
~{\rm Tr}(
\sqrt{-g}
 A_t \Pi^{k t} 
 ).
 \end{eqnarray}

The electric charge can also be expressed as a surface integral,
a natural definiton of it being
\begin{eqnarray}
\label{e-charge}
Q_e=\frac{1}{4\pi}\oint_{\infty}dS_{k}
{\rm Tr}( \hat{\Phi}\Pi^{k t} ),
\end{eqnarray}
for all  YMH models to be discussed below 
(where we define as usual $\hat{\Phi}=\Phi/|\Phi|$).

The definition of the magnetic charge is dependent of the particular model, and these magnetic
monopole charge densities will be defined for each model in turn, below.

\section{The specific models}

\subsection{The  Georgi-Glashow model on $\R^{3,1}$: the $p=1$ model}
The 'canonical' model of a YMH theory is the 
 GG one, with a Lagrangian\footnote{This Lagrangian
is usually supplemented with a Higgs potential,
$V(\Phi)=\frac{\lambda}{4} (|\Phi^a|^2 - \eta^2)^2$.
However, to simplify the general picture,
we shall not consider this term which is not of interest for the purposes of this work.
}
\be
\label{lagYMH1}
{\cal L}^{(1)}=-\frac{1}{4} \mbox{Tr} \bigg( F_{\mu\nu}\,F^{\mu\nu} \bigg )
-\frac12 \mbox{Tr} \bigg ( D_{\mu}\F\,D^{\mu}  \Phi \bigg ),
\ee
in which case
\be
\label{suo1}
\Pi^{\mu \nu}=F^{\mu\nu},~~\Pi^\mu=D^{\mu}  \Phi.
\ee
%
The stress tensor of this model, as read from (\ref{canonical-Tik-1}) is (here we give the (more transparent)
expression in terms of $F_{\mu\nu}$ and $D_{\mu}  \Phi$)
\be
\label{strYMH1}
2\,T_{\mu\nu}^{(1)}= \mbox{Tr}  \left(F_{\mu\tau}\,F_{\nu}{}^{\tau}-\frac14\,g_{\mu\nu}\,F_{\tau\la}\,F^{\tau\la}\right)
- \mbox{Tr}\,\left(D_{\mu}\F\,D_{\nu}\F-\frac12\,g_{\mu\nu}\,D_{\tau}\F\,D^{\tau}\F\right).
\ee

The GG model has a variety of interesting features which have been extensively studied in the literature over the last 40 years.
Here we mention only that 
in the static, purely magnetic limit ($i.e.$ $A_t=0$), the  Hamiltonian of the model (which 
in that case coincide with the Lagrangian) 
is bounded from below by the magnetic monopole charge density
\be
\label{1}
\varrho^{(1)}=\frac{1}{4\pi}\,\vep_{ijk}\,\mbox{Tr}\,(F_{ij}D_k\F) \stackrel{{\rm def.}}=\bnabla\cdot\bOmega^{(1)},~~
{\rm with}~~
\bOmega^{(1)}=\frac{1}{4\pi}\vep_{ijk}\,\mbox{Tr}\,(\F\,F_{ij}),
\ee
which is the dimensional descendant of the $2$-nd Chern-Pontryagin density. 
The total magnetic charge $Q_m$ is the integral of $\varrho^{(1)}$.

\subsection{The $(p=1)$+$(p=2)$ model}

 In our choice of this  YMH model, we 
start from the simple observation that the GG model (\ref{lagYMH1})
can be obtained as descending from the pure $F^2$ YM Lagrangian on $R^3\times S^1$,
the components of the YM potential along the $S^1$ direction corresponding to the Higgs fields.
This observation has led in \cite{Tch}
to the construction of a hierarchy of $p\geq 1$  
$SO(3)$ Higgs models in $D=3$ space dimensions, 
obtained by dimensional descent over $S^{4p-3}$, of the $p^{th}$ member of the Yang-Mills hierarchy.
The $p = 1$ member here is nothing else
than the Prasad-Sommerfield (PS) limit of the GG model.
For any $p\geq 2$, the Lagrangian ${\cal L}^{(p)}$ 
is a Skyrme-like gauged Higgs system,
in the sense that it consists of $2p^{th}$ powers
of the gauge curvature $F(2)$ 
and covariant derivative $D\F$, suitably antisymmetrised so that only the
squares of `velocity'-fields appear.
Moreover, for each $p\geq 1$ YMH model, one can define a topological
charge density which is the descendant of the $2p^{th}$ Chern-Pontryagin density. In each case there exists a
Bogolmol'nyi bound, but this bound can be saturated only in the $p=1$ case.

 In the present work, we restrict our attention to the system described by
the Lagrangian of the  $(p=1)$+$(p=2)$ model, which consists of the sum of the 
 GG model in the PS limit ($i.e.$ $p=1$) plus a ``correction part'',
  which is inspired by the Lagrangian of the 
  $p=2$ term in the general YMH hierarchy introduced in \cite{Tch}
\begin{eqnarray}
\label{gen-L}
{\cal L}={\cal L}^{(1)}
+
{\cal L}^{(2)}.
\end{eqnarray}
It is clear that restricting to ``corrections'' of the $p=2$ YMH is sufficient to describe
the effect of such correction qualitatively. The $p=2$ YMH model is described in detail in the next subsection.

\subsubsection{The $p=2$ YMH system on $\R^{3,1}$}
 
The Lagrangian of the general $p=2$ model can written as a sum of five different terms
\be
\label{LYMH2}
\frac{1}{20}{\cal L}^{(2)}=\sum_{a=0}^4\lambda_a {\cal L}^{(2,a)} ,
\ee
with
\begin{eqnarray}
{\cal L}^{(2,0)} =-\frac14\ {\cal F}_{\mu\nu\rho\si}{\cal F}^{\mu\nu\rho\si},~~
{\cal L}^{(2,1)} ={\cal F}_{\mu\nu\rho}{\cal F}^{\mu\nu\rho},~~
{\cal L}^{(2,2)} =6 {\cal F}_{\mu\nu}{\cal F}^{\mu\nu},~~
{\cal L}^{(2,3)} =-9 {\cal F}_{\mu}{\cal F}^{\mu},~~
{\cal L}^{(2,4)} =-54 {\cal F}^2,
\end{eqnarray}
where we have used the symbolic notation stated in (\ref{cal2}).

The Lagrangian of the original $p=2$ YMH model (obtained by dimensional descent from a pure $F^4$ Yang-Mills model in $D=8$ dimensions)
is found by taking 
$\lambda_0=0$, $\lambda_1=\lambda_2=\lambda_3=\lambda_4=1$.
However, it is of interest to treat $\lambda_a$ as arbitrary positive parameters\footnote{In what follows, we take $\lambda_a\geq 0$
as implied by requiring a strictly positive energy density.} 
and to work with the generic model (\ref{LYMH2}). 
Moreover, the term ${\cal F}_{\mu\nu\rho\si}{\cal F}^{\mu\nu\rho\si}$ has been added by hand in (\ref{LYMH2}) since it does not occur in the
original $p=2$ YMH model.
Note that in the purely magnetic limit, this term is absent, and therefore it is interesting to see what effect its inclusion has
in the construction of the electrically charged solutions\footnote{As discussed in \cite{Radu:2011ip},
a ${\cal F}_{\mu\nu\rho\si}{\cal F}^{\mu\nu\rho\si}$ 
term stabilizes the EYM hairy black holes,
leading to solutions with very different properties
as compared to the standard ones \cite{Volkov:1998cc}.
Moreover, this term leads to new qualitative features in inflationary models
with non-Abelian gauge fields \cite{Maleknejad:2011jw}.
}.

The  expressions of the generalized momentum densities 
for each of the terms in \re{LYMH2} is
\bea
\Pi^{\mu\nu}_{(2,0)}=-3!\,\{{\cal F}^{\mu\nu\rho\si},F_{\rho\si}\},
~~
\Pi^{\mu\nu}_{(2,1)}=2\cdot\, 3!\{{\cal F}^{\mu\nu\rho},D_{\rho}\F\},
~~
\Pi^{\mu\nu}_{(2,2)}=-4!\,\{S,{\cal F}^{\mu\nu}\},
\label{mn2} 
\eea
and
\bea
\Pi^{\mu}_{(2,1)}=3!\,\{{\cal F}^{\mu\rho\si},F_{\rho\si}\},
~~
\Pi^{\mu}_{(2,2)}=-4![{\cal F}^{\mu\rho},D_{\rho}\F],
~~
\Pi^{\mu}_{(2,3)}=-18\,\{S,{\cal F}^{\mu}\}.
\label{m3}
\eea
The corresponding symmetric and gauge invariant energy-momentum tensor reads
(here we give the final expression used in practice, in terms of $F_{\mu\nu}$ and $D_\mu \Phi$):
\bea
\label{strYMH2}
\frac{1}{20}T_{\mu\nu}^{(2)}&=&-\la_0\,\mbox{Tr}\,
\left(F_{\mu\tau\la\si}\,F_{\nu}{}^{\tau\la\si}-\frac18\,g_{\mu\nu}\,F_{\ka\tau\la\si}\,F^{\ka\tau\la\si}\right)
\nonumber\\
&+&3\la_1\,\mbox{Tr}\,\left(\left\{F_{[\mu\tau},D_{\la]}\F\right\}\,\left\{F_{[\nu}{}^{\tau},D^{\la]}\F\right\}
-\frac16\,g_{\mu\nu}\,\left\{F_{[\mu\nu},D_{\rho]}\F\right\}\,\left\{F^{[\mu\nu},D^{\rho]}\F\right\}\right)\nonumber\\
&+&2\cdot6\,\la_2\,\mbox{Tr}\,\bigg(\left(\{S,F_{\mu\tau}\}+[D_{\mu}\F,D_{\tau}\F]\right)\,\left(\{S,F_{\nu}{}^{\tau}\}
+[D_{\nu}\F,D^{\tau}\F]\right)\nonumber\\
&&\qquad\qquad-\frac14\,g_{\mu\nu}\,\left(\{S,F_{\tau\la}\}+
[D_{\tau}\F,D_{\la}\F]\right)\,\left(\{S,F^{\tau\la}\}+[D^{\tau}\F,D^{\la}\F]\right)
\bigg)\nonumber\\
&-&9\,\la_3\,\mbox{Tr}\,\left(\{S,D_{\mu}\F\}\{S,D_{\nu}\F\}-\frac12\,g_{\mu\nu}\{S,D_{\la}\F\}\{S,D^{\la}\F\}\right)\nonumber\\
&-&54\,\la_4\,\mbox{Tr}\left(0-\frac12\,g_{\mu\nu}\,S^4\right)\, .
\eea
It may be interesting to display the topological charge density of the $p=2$ monopole, alluded to in the previous subsection.
\bea
\varrho^{(2)}&\simeq&\vep_{ijk}\,\mbox{Tr}
\bigg(3\eta^4F_{ij}D_k\F+\eta^2\left[3F_{ij}\left(\F^2D_k\F+D_k\F\F^2\right)-2D_i\F D_j\F D_k\F\right]\nonumber\\
&&\qquad\qquad\quad+\left[F_{ij}\left(\F^4D_k\F+D_k\F \F^4+\F^2D_k\F\F^2\right)-2\F^2D_i\F D_j\F D_k\F\right]\bigg)
\label{2}\\
&\stackrel{{\rm def.}}=&\bnabla\cdot\bOmega^{(2)}.
\nonumber
\eea
This pertains to the system \re{LYMH2} (with $\lambda_0=0$, 
$\lambda_1=\lambda_2=\lambda_3= \lambda_4=1$) on $\R^3$ and
descends from the $4^{th}$ Chern-Pontryagin density in $8$ dimensions.
It is manifestly a total divergence
\bea
\Omega^{(2)}_k \simeq \vep_{ijk}\,\mbox{Tr}\bigg[3\eta^4\F F_{ij}+2\eta^2
\left(\F^3 F_{ij}-\F D_i\F D_j\F\right) 
 +\frac15\left[3\F^5F_{ij}-2\left(2\F^3D_i\F D_j\F-\F^2D_i\F\,\F\,D_j\F\right)\right]
\bigg].
\label{ttd2}
\eea
It is interesting to note that the surface integrals of \re{ttd2} and \re{1} are equal, up to a numerical multiple. This
is because the terms in \re{ttd2} not featuring the curvature $F_{ij}$ decay too fast to contribute, by virtue of Higgs
asymptotics. This density was employed in \cite{Kleihaus:1998kd,Kleihaus:1998gy}.


\subsection{The non-Abelian  Born-Infeld--Higgs model}
Another interesting possibility is to
consider a  Born-Infeld (BI) Lagrangian for the 
gauge fields.
This modification of the standard YM quadratic Lagrangian
is suggested by the superstring theory 
\cite{Tseytlin:1997csa},
\cite{Gauntlett:1997ss}
leading to a variety of interesting features.
For example, the non-go results in \cite{Deser:1976wq}
forbidding the existence of pure YM solitons
are circumvented in this case, since
a  non-Abelian BI theory possesses
particle-like solutions even in the absence of a Higgs field \cite{Gal'tsov:1999vn}.
The spherically symmetric monopoles and dyons of this model
have been studied in 
\cite{Grandi:1999rv},
\cite{Grandi:1999dv}.

The Lagrangian of the non-Abelian  BI--Higgs theory reads
\begin{eqnarray}
\label{LBIH}
{\cal L}^{(BIH)} = 
\beta^2 \bigg(
1-\sqrt
{1+U}~
\bigg)
-\frac12 \mbox{Tr} \bigg ( D_{\mu}\F\,D^{\mu}  \Phi \bigg )
,~~
{\rm with}~~ 
U=\frac{1}{2\beta^2} \mbox{Tr}\big ( F_{\mu\nu}\,F^{\mu\nu} \big )
-\frac{1}{\beta^4} \mbox{Tr}\big ( F_{\mu\nu \rho\sigma}\,F^{\mu\nu\rho\sigma} \big )~,
\end{eqnarray}
where \mbox{Tr} represents the usual trace on $SO(3)$ 
indices. 
(Note that the definition of the BI theory for a non-Abelian gauge group is not 
unique and several alternatives have been
discussed in the literature.
Apart from the one used above, another possibility
of interest (which will shall not consider here) 
is to take a symmetric trace operation \cite{Tseytlin:1997csa},
but so far the explicit Lagrangian with such trace is
known only as perturbative series.)

The corresponding expressions of the generalized momentum densities read
\begin{eqnarray}
\label{P-BI}
\Pi^{\mu\nu}=-\frac{1}{\sqrt{1+U}}
\bigg(
F^{\mu\nu}
-\frac{2}{\beta^2} 
\,\{{\cal F}^{\mu\nu\rho\si},F_{\rho\si}\}
\bigg),~~
\Pi^{\mu}=D^{\mu}\Phi.
\end{eqnarray} 
In the absence of a topological lower bound, it is natural to use  
the magnetic charge definition valid for the GG model, 
also in this case.

\subsection{Yang-Mills--Higgs model with Chern-Simons--like term}
 In $2+1$ dimensions, electric charge and angular momentum result~\cite{Paul:1986ix,Hong:1990yh,Jackiw:1990aw}
from the dynamics of a Chern-Simons term in the Lagrangian. In $3+1$ dimensions however, no Chern-Simons density was
identified until recently proposed in \cite{Tchrakian:2010ar,Radu:2011zy}. In a given spacetime dimension, they result from the descent by one step, of the $1$-form
in the total divergence expression of a Chern-Pontryagin (CP) density, which itself is a dimensional descendant of a CP density in some higher (even)
dimension. These 'new' Chern-Simons densities feature both gauge fields and Higgs fields,
and are defined in both odd and even spacetime dimensions.

Here, we consider the simplest example in $3+1$ dimensional spacetime,
which was discussed recently in \cite{Navarro-Lerida:2013pua}. 
This CS density is extracted from the dimensionally reduced
CP density on ${\bf M}^5\times{S^1}$. 
The residual CP density on ${\bf M}^5$ being a total divergence, a Chern-Simons density in $3+1$ dimensions
can be extracted in the usual way. The residual gauge group then is $SO(5)$ and the Higgs field takes its values in the algebra of $SO(6)$.

The Lagrangian density of this specific Yang-Mills--Higgs--Chern-Simons (YMHCS) model
reads
\begin{eqnarray}
\label{L-YMHCS}
{\cal L} =-\frac{1}{4} \mbox{Tr} \bigg( F_{\mu\nu}\,F^{\mu\nu} \bigg )
-\frac12 \mbox{Tr} \bigg ( D_{\mu}\F\,D^{\mu}  \Phi \bigg )
+i \kappa \epsilon^{\mu\nu\rho\sigma} 
\mbox{Tr}  \big ( \Phi F_{\mu\nu} F_{\rho\sigma} \big ),
\end{eqnarray}
with $\kappa$ an arbitrary constant.
The generalized momentum densities are
\begin{eqnarray}
\label{P-YMHCS}
\Pi^{\mu\nu}= 
F^{\mu\nu}+ i \kappa \epsilon^{\mu\nu\rho\sigma} 
\mbox{Tr}  \big ( \Phi F_{\rho\sigma} \big ),~~
\Pi^{\mu}=D^{\mu}\Phi.
\end{eqnarray}
Following \cite{Navarro-Lerida:2013pua},  
 we shall restrict our study to Yang-Mills
 fields taking their values in the $SO(3)\times U(1)$ subalgebra of $SO(5)$, with an $SO(3)$  Higgs triplet. 
In this limit, the Lagrangian (\ref{L-YMHCS}) of the model can be written in a equivalent form as
 \begin{eqnarray}
 \label{GKLTT}
{\cal L} =
-\frac{1}{4} \mbox{Tr} \bigg( F_{\mu\nu}\,F^{\mu\nu} \bigg )
-\frac12 \mbox{Tr} \bigg ( D_{\mu}\F\,D^{\mu}  \Phi \bigg )
-\frac{1}{4} f_{\mu\nu}f ^{\mu\nu}  
+i\kappa \epsilon^{\mu\nu\rho\sigma} 
 f_{\mu\nu}
\mbox{Tr}  \big ( \Phi F_{\rho\sigma} \big ),
\end{eqnarray}
with $f_{\mu \nu}=\partial_{\mu}a_{\nu}-\partial_{\nu}a_{\mu}$
a $U(1)$ field and $F_{\mu\nu}$, $\Phi$ the usual SO(3) gauge fields and Higgs field, respectively.
As found in \cite{Navarro-Lerida:2013pua}, this extra term prevents the existence of (self-dual) solutions saturating a lower bound.

The Lagrangian (\ref{L-YMHCS}) corresponds to the flat spacetime limit of a
YMH-Maxwell model considered in Section {\bf 2} of \cite{Gibbons:1993xt} in a more general context.
Interestingly, the results in \cite{Gibbons:1993xt} show that when including the gravity effect, the self-gravitating solitonic solutions of (\ref{GKLTT})
saturate a gravitational version of the Bogomol'nyi bound, with the same magnetic charge definition as in the GG model.

\section{The issue of spinning solutions }

\subsection{Axially symmetric YMH fields}
In order to evaluate the general relation (\ref{J}) and to construct
generalized dyons and dipole, we need to specific a YMH ansatz.
 Employing spherical coordinates, we use the following parametrization of 
the axially symmetric YMH fields:
\begin{eqnarray}
&&A_\mu dx^\mu
=\left( H_5 \frac{i\tau_r^{(n)}}{2}+ H_6 \frac{i\tau_\theta^{(n )}}{2 } \right)dt+
\left[ \frac{H_1}{r} dr + (1-H_2)d\theta\right] \frac{i\tau_\varphi^{(n)}}{2 }
- n \sin\theta \left( H_3\frac{i\tau_r^{(n )}}{2 }
                     + H_4\frac{i\tau_\theta^{(n )}}{2 }\right) d\varphi
\ ,  \, \, \, \, \, \, \, \, \, \, \, \, \label{ansatzA}
\\
&&
\Phi
=
 \eta \left( \Phi_1\frac{i \tau_r^{(n )}}{2}+ \Phi_2\frac{i\tau_\theta^{(n )}}{2}  \right),
\label{ansatzPhi}
\end{eqnarray}
in terms of six gauge potentials and two Higgs functions.
The only $\varphi$-dependent terms in (\ref{ansatzA}) are the $SU(2)$ matrices
$\tau_r^{(n )} $, $\tau_\theta^{(n )} $  and $\tau_\varphi^{(n)}$.
These matrices 
are defined as  
$
\tau_r^{(n )} =
\sin  \theta  \left(\cos n\varphi~\tau_x + \sin n\varphi~\tau_y \right) + \cos \theta~\tau_z \ ,
$
$
\tau_\theta^{(n )} =
\cos \theta  \left(\cos n\varphi~\tau_x + \sin n\varphi~\tau_y \right) - \sin  \theta~\tau_z \ ,
$
and
$
\tau_\varphi^{(n)} =
 -\sin n\varphi~\tau_x + \cos n\varphi~\tau_y \ ,
 $
(with $\tau_a = (\tau_x, \tau_y, \tau_z)$ the Pauli matrices).
For $H_5=H_6=0$, (\ref{ansatzA})  is essentially the  axially symmetric YMH ansatz as introduced 
  by Manton \cite{7}, and Rebbi and Rossi \cite{Rebbi:1980yi}  when discussing multimonopole solutions. 
This particular parametrisation of the YM ansatz in terms of $H_i$ is very
convenient for numerical studies
and is employed in most of the work on axially symmetric YMH systems.
Note that (\ref{ansatzPhi}) contains, via the matrices
$\tau_r^{(n )} $, $\tau_\theta^{(n )} $  and $\tau_\varphi^{(n)}$, an extra integer $n=1,2,\dots$
which is the winding number of the solutions. 

Let us also mention that the above Ansatz
possesses a residual Abelian gauge
invariance (see $e.g.$ the discussion in \cite{Brihaye:1994ib}).
To fix it, we 
have to include a gauge fixing term, the most convenient choice being
$
r \partial_r H_1 - \partial_\theta H_2 = 0.
$
 
\subsection{The far field asymptotics}
For any choice of the Lagrangian,
the assumption that the Higgs field approaches
asymptotically a constant value $|\Phi| \to 1$  
(which is the effect of a symmetry breaking Higgs potential,
whether explicitly included in the action or not),
 results in the following finite energy condition on the Higgs
 \begin{eqnarray}
 \label{Higgs-asympt}
\Phi^1=\cos (m-1)\theta, ~~\Phi^2=\sin (m-1)\theta,
 \end{eqnarray}
 with $m=1,2,\dots$ a positive integer.
 In the next step, we shall assume that the gauge derivative of the Higgs fields
 vanished asymptotically,
 $D_{\mu}\Phi D_{\mu}\Phi \to 0$,
a condition which then fixes the asymptotic values of the YM potentials.
 For odd $m$ these are
\begin{eqnarray}
\label{oddm}  
&&
H_1=0,~~H_2=-(m-1),~~
H_3=\frac{\cos \theta }{ \sin \theta}[\cos(m-1)\theta -1],~~
H_4=-\frac{\cos \theta}{\sin \theta}\sin(m-1)\theta,
\\
\nonumber
&&
H_5= V_0 \cos (m-1)\theta,~~H_6= V_0 \sin (m-1)\theta,~
\end{eqnarray}
and for even $m$
\begin{eqnarray}
\label{evenm}
&&
H_1=0,~~H_2=-(m-1),~~
H_3=\frac{1}{\sin \theta}[\cos(m-1) \theta  -\cos \theta ],~~
H_4=-\frac{  \sin(m-1)\theta }{\sin \theta},
\\
\nonumber
&&
H_5= V_0 \cos (m-1)\theta,~~H_6= V_0 \sin (m-1)\theta,~
\end{eqnarray}
 with $V_0$ a constant.
(A derivation of these boundary conditions  can be found in  
\cite{Radu:2004ys}, 
\cite{Kleihaus:2003nj}, 
\cite{Kleihaus:2005fs}.)
 Thus,
for $m=2k$ ($k=1,2,\dots$),
the ground state of the model  corresponds to
a gauge transformed trivial solution and the magnetic charge vanishes.
 The situation is different for odd values $m=2k+1=1,3,\dots$,
the ground state in this case corresponding to a charge $n$   multimonopole.

To evaluate (\ref{J}),  we need both $W$, which for the axially symmetric configuration at hand is
\be
\label{W1}
W=-n(\cos \theta + \sin \theta H_3)\frac{i\tau_r^{(n)}}{2}
+n \sin \theta H_4\frac{i\tau_\theta^{(n)}}{2} \,,
\ee
and, the asymptotic expression of the generalized momentum $\Pi^{rt}$.
At this point, we remark that, for all models in this work, the YM Lagrangian effectively includes the quadratic term $F_{\mu\nu}^2$,
namely the $p=1$ YM term. Now the additional curvature terms  YM terms are all
higher order in the curvature $2$-form, so they decay faster than the usual $p=1$, 
and hence their contributions to the \re{J} integral will vanish. 
Then the following expression holds for large $r$: $\Pi^{r t}=F^{rt}+({\rm subleading~terms})$,
and the electric potentials
$H_5$ and $H_6$
have long-range, Coulomb-like tails,
 \begin{eqnarray}
 \label{H56Q}
H_5=\cos (m-1)\theta~(V_0 -\frac{Q_e}{r})+\dots,~~
H_6=\sin (m-1)\theta~(V_0 -\frac{Q_e}{r})+\dots~,
 \end{eqnarray}
 with $Q_e$ corresponding to the electric charge as computed from (\ref{e-charge}). 
 
We close this part by noticing that the {\it `electric mass'} (\ref{E}) can
always be written as a product between the `electrostatic potential' $V_0$ 
and the electric charge $Q_e$,
 \begin{eqnarray}
 \label{E1}
 E_e=\frac{1}{2}V_0 Q_e.
  \end{eqnarray}
Assuming that the {\it 'electric'}
part of the energy density ${\cal E}$, as given by (\ref{electric1}), is a positive quantity,
it follows that a configuration with 
$V_0=0$
or $Q_e=0$
does not possess an electric field $F_{k t}\equiv 0$.

Another important feature is that the `electrostatic potential' $V_0$ 
(as fixed by the asymptotics of the functions $H_5$, $H_6$)
is always bounded from above by the $v.e.v.$ of the Higgs
field ($i.e.$ $V_0<1$ for the conventions here).
Since the $p=1$ term
would dominate as $r\to \infty$,  
a value $V_0>1$
 implies a far field oscillatory behaviour of some YM potentials,
a feature which is not compatible with finite energy requirements
(this can be seen explicitly in the Prasad-Sommerfield
exact solution
\cite{Prasad:1975kr};
see also
\cite{Hartmann:2000ja} 
for the axially symmetric case).

\subsection{The total angular momentum }
With these relations at hand, the evaluation of the general relation (\ref{J}) is straightforward. 
For the parametrization (\ref{ansatzA}), the asymptotic expression of $W$ which enters the general relation  
(\ref{relations}) is  
\[
W=-n \cos \theta \left(\cos(m-1)\theta  \frac{i\tau_r^{(n)}}{2}+\sin(m-1)\theta  \frac{i\tau_\theta^{(n)}}{2} \right) ,
\]
for odd $m$
and
\[
W=-n  \left( \cos(m-1)\theta  \frac{i\tau_r^{(n)}}{2}+\sin(m-1)\theta  \frac{i\tau_\theta^{(n)}}{2}\right)  ,
\]
for even $m$.
Also, for any $m$, the asympototic expression of the generalized momentum $\Pi^{rt}$
is
\bea
\label{Pirt}
\Pi^{rt}= \left(\cos (m-1)\theta~ \frac{i \tau_r^{(n )}}{2}
+
\sin (m-1)\theta~ \frac{i \tau_\theta^{(n )}}{2} \right)\frac{Q}{r^2}
+\dots~.
\eea
Then, after replacing in (\ref{J}), 
one finds that for both $(p=1)+(p=2)$ models and
the BI-Higgs models
the following relations hold
\bea
\label{JQ}
&&
J=0~~~~~~{\rm for}~~m=1,3,\dots~~(Q_M\neq 0), 
\\
\nonumber 
&&
J=n Q_e~~{\rm for}~~m=2,4,\dots~~(Q_M= 0).
\eea
This is the relation derived before for the pure GG case,
connecting  the quantization of charge and angular momentum.
However, we see that this is a more general result, which holds as long as
the YMH fields share the asymptotics of the GG model (while the precise features in the bulk can be very different).
Therefore we conclude that a magnetic monopole does not spin while a magnetic dipole cannot rotate
unless it is endowed with a net electric charge. 

The situation is more involved for the model (\ref{GKLTT}) featuring a Chern-Simons like term,
since the angular momentum density has a supplementary part originating from this extra term.
To evaluate this supplementary contribution, we write the usual axially symmetric
Ansatz for the U(1) field $1$-form
 \bea
a=a_\varphi d\varphi+a_t dt.
\eea
The boundary condition by $a_\mu$ as $r\to \infty$ are $a_\varphi \to 0$ and $a_t\to v_0$ (with $v_0$
an arbitrary constant). 
Moreover, one can easily see that the asymptotic behaviour of the YMH field remains the same as in the GG model.
Then a straightforward computation shows that the general relations (\ref{JQ}) still holds for this model.
Thus we conclude that the supplementary YMH CS-like term in (\ref{GKLTT}) 
does not endow a monopole with a  nonzero angular momentum.

 However, one should remark
that the charge-parity violating term in \re{GKLTT} is not strictly speaking a Chern-Simons term, unlike the corresponding term in \re{L-YMHCS}.
Thus, any statements concerning the vanishing of the angular momentum is not $apriori$ valid for the general Chern-Simons theory \re{L-YMHCS}.
It is concievable that in that case the angular momentom may not vanish.
Any progress in this direction requires first an investigation of the issue of spinning
monopoles
for higher gauge groups and for different representations of the Higgs field,
a task which has not been yet considered in the literature.

\section{Generalized dyons and magnetic dipoles in 
$(p=1)+(p=2)$ YMH model. Numerical results}
Although we have seen that all solutions share the same relation 
between the angular momentum and electric and magnetic charges
as in the GG model, 
 it remains
an interesting question to see how  
a more general YMH Lagrangian
{\it quantitatively} affects the properties of the known axially symmetric 
configurations.

To answer this question, we have 
considered the $(p=1)+(p=2)$ YMH model
and constructed generalizations of the simplest  solutions
corresponding to dyons and  magnetic dipoles.
To our knowledge, this problem has not been addressed before in the literature,
only purely magnetic solutions being considered so far.
The monopole solutions of the $(p=1)+(p=2)$ model have been discussed in  
 \cite{Kleihaus:1998kd},
 \cite{Kleihaus:1998gy},
 a rather complicated picture being revealed.
 For example,  this model does not support selfdual
solutions 
\cite{Tchrakian:1990gc}.
Moreover, depending on the values of the dimensionless coupling parameters,
the generalised monopoles exhibit
both attractive and repulsive phases.

In this Section we extend the results of 
\cite{Kleihaus:1998kd} in several different 
directions, by including an electric component in the YM connection.

\subsection{The boundary conditions}
The   boundary conditions at infinity for dyons are found by taking $m=1$
 in the general asymptotics (\ref{Higgs-asympt}), (\ref{oddm})
\begin{eqnarray}
\label{MM-inf}
H_1|_{r=\infty}=H_2|_{r=\infty}=H_3|_{r=\infty}=H_4|_{r=\infty}=H_6|_{r=\infty}=\Phi_2=0,~~
H_5|_{r=\infty}=V_0,~~\Phi_1=1,
\end{eqnarray}
with $V_0<1$ a constant corresponding to the electric potential.
 The corresponding boundary conditions at the origin $r=0$ are
\begin{eqnarray}
\label{MM-r0}
H_1|_{r=0}=H_3|_{r=0}=H_5|_{r=0}=H_6|_{r=0}=\Phi_1|_{r=0}=\Phi_2|_{r=0}=0,~~
 H_2|_{r=0}=H_4|_{r=0}=1.
\end{eqnarray}
On the symmetry axis, the dyons satisfy the
boundary conditions
\begin{eqnarray}
\label{MM-z}
H_1|_{\theta=0,\pi}= H_3|_{\theta=0,\pi}= H_6|_{\theta=0,\pi}=\Phi_2|_{\theta=0,\pi}=0,~~
\partial_\theta H_2|_{\theta=0,\pi}=\partial_\theta H_4|_{\theta=0,\pi}=\partial_\theta \Phi_1|_{\theta=0,\pi}=0.
\end{eqnarray}

However,
as discussed for the first time in \cite{Taubes:1982ie} for the GG model, 
there
exist also a different type of solutions of the second order
Euler-Lagrange equations, which are not stable and represent
saddle points of the energy, rather than absolute minima.
In the absence of an electric field, they correspond to  magnetic dipoles.
A systematic discussion of the properties of these solution in GG model is given in ref.
\cite{Kleihaus:1999sx}
(note that these are non-BPS configurations and no exact solution is known in this case).
For example, the magnetic charge  
measured at infinity vanishes, despite the existence locally of a nonzero density. 
This, in the presence of an electric charge, results in nonzero angular
momentum.
The electrically charged, spinning version of the dipole solutions 
are studied in
\cite{Paturyan:2004ps},
\cite{Kleihaus:2005fs}.

The magnetic dipole solutions satisfy a different set of boundary conditions at infinity
 than 
(\ref{MM-inf}), 
(\ref{MM-r0}), 
(\ref{MM-z}).
These boundary conditions  are found by taking $m=2$ 
in the  general expressions (\ref{Higgs-asympt}), (\ref{evenm})
\begin{eqnarray}
\nonumber
&&
H_1|_{r=\infty}=H_3|_{r=\infty}=0,~H_2|_{r=\infty}=H_4|_{r=\infty}=-1, 
\\
\label{MA-inf}
&&
H_5|_{r=\infty}=V_0 \cos \theta,~~H_6|_{r=\infty}=V_0 \sin \theta,~~
\Phi_1|_{r=\infty}= \cos \theta,~~\Phi_2|_{r=\infty} =\sin \theta,~~
\end{eqnarray}
 (with $V_0<1$ again).
The other boundary conditions are
\begin{eqnarray}
 \nonumber
&&
H_1|_{r=0}=H_3|_{r=0}=0,~
 H_2|_{r=0}=H_4|_{r=0}=1,~
 \\
\label{MA-r0}
&& (\cos \theta \partial_r H_5- \sin \theta \partial_r H_6)|_{r=0}=0,~
(\sin \theta H_5+\cos \theta H_6)|_{r=0}=0,~~
\\
\nonumber
&&(\cos \theta \partial_r \Phi_1- \sin \theta \partial_r \Phi_2)|_{r=0}=0,~
(\sin \theta\Phi_1+\cos \theta \Phi_2)|_{r=0}=0,
\end{eqnarray}
at the origin, and 
\begin{eqnarray}
\label{MA-z}
H_1|_{\theta=0,\pi}= H_3|_{\theta=0,\pi}= H_6|_{\theta=0,\pi}=\Phi_2|_{\theta=0,\pi}=0,~~
\partial_\theta H_2|_{\theta=0,\pi}=\partial_\theta H_4|_{\theta=0,\pi}=\partial_\theta \Phi_1|_{\theta=0,\pi}=0,
\end{eqnarray}
on the symmetry axis.

\subsubsection{$n=1$ results: spherically symmetric generalized dyons}
The spherically symmetric solutions are found by taking
\begin{eqnarray}
\label{sph1}
H_1=H_3=H_6=\Phi_2=0,~~
H_2=H_4=w(r),~~H_5=u(r),~~\Phi_1=h(r),
\end{eqnarray}
in the general axially symmetric parametrization 
(\ref{ansatzA})-(\ref{ansatzPhi}), and have a winding number $n=1$.
The boundary conditions in this case can be read from (\ref{MM-inf}) together with (\ref{MM-r0}).

The one dimensional reduced Lagrangian of this system is given by the sum of the $p=1$ and $p=2$ terms,
\be
\label{redsphlag1}
{\cal L}_{\rm YMH}^{(1)}=-\frac14\,r^2\,\left[2\left(\frac{w'}{r}\right)^2+\frac{1}{r^4}(1-w^2)^2\right]
-r^2\left[(\eta^2\,h'^2-u'^2)+2\,\left(\frac{w}{r}\right)^2\,(\eta^2\,h^2-u^2)\right],
\ee
and
\bea
{L}_{\rm YMH}^{(2)}&=&6 \left\{\la_0\,\left([(1-w^2)\,u]'\right)^2-\la_1\,\eta^2\left([(1-w^2)\,h]'\right)^2\right\}
\nonumber
\\
\label{redsphlag21}
&-&12\,\la_2\,\eta^4 \left\{2\left([(1-h^2)\,w]'\right)^2+r^{-2}\left([(1-w^2)(1-h^2)+2w^2h^2]\right)^2\right\}
\\
&+&12\,\eta^4 r^2\,(1-h^2)^2\left\{4\,\la_2\,[u'^2+2r^{-2}w^2u^2]-3\,\la_3\,\eta^2\,[h'^2+2r^{-2}w^2h^2]\right\}
\nonumber
\\
\nonumber
&-&54\,\la_4\,\eta^8\,r^2(1-h^2)^4.
\eea
The first of these, \re{redsphlag1}, supports the usual Julia-Zee dyon, while the second one, \re{redsphlag21}
 supports the next excited Julia-Zee dyon. Here, we have
analysed the dyon solutions of a combination of these systems.

The system of three non-linear coupled differential equations
for the functions  $w,h$ and $u$, subject to the  boundary conditions described above,
was solved  by using the software package COLSYS developed
by Ascher, Christiansen and Russell \cite{COLSYS}.
This solver employs a collocation
method for boundary-value ordinary differential equations and a damped Newton method
of quasi-linearization. 

\newpage
\setlength{\unitlength}{1cm}
\begin{picture}(8,6)
\put(-0.5,0.0){\epsfig{file=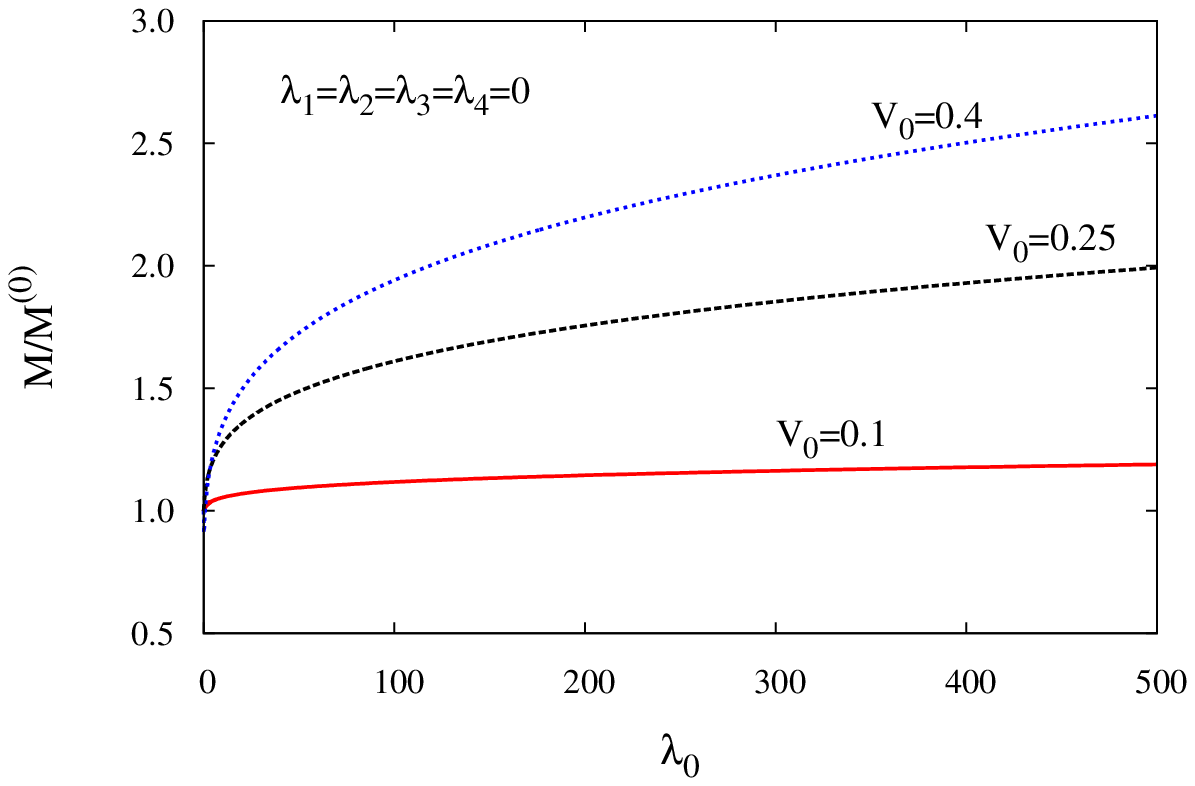,width=8cm}}
\put(8,0.0){\epsfig{file=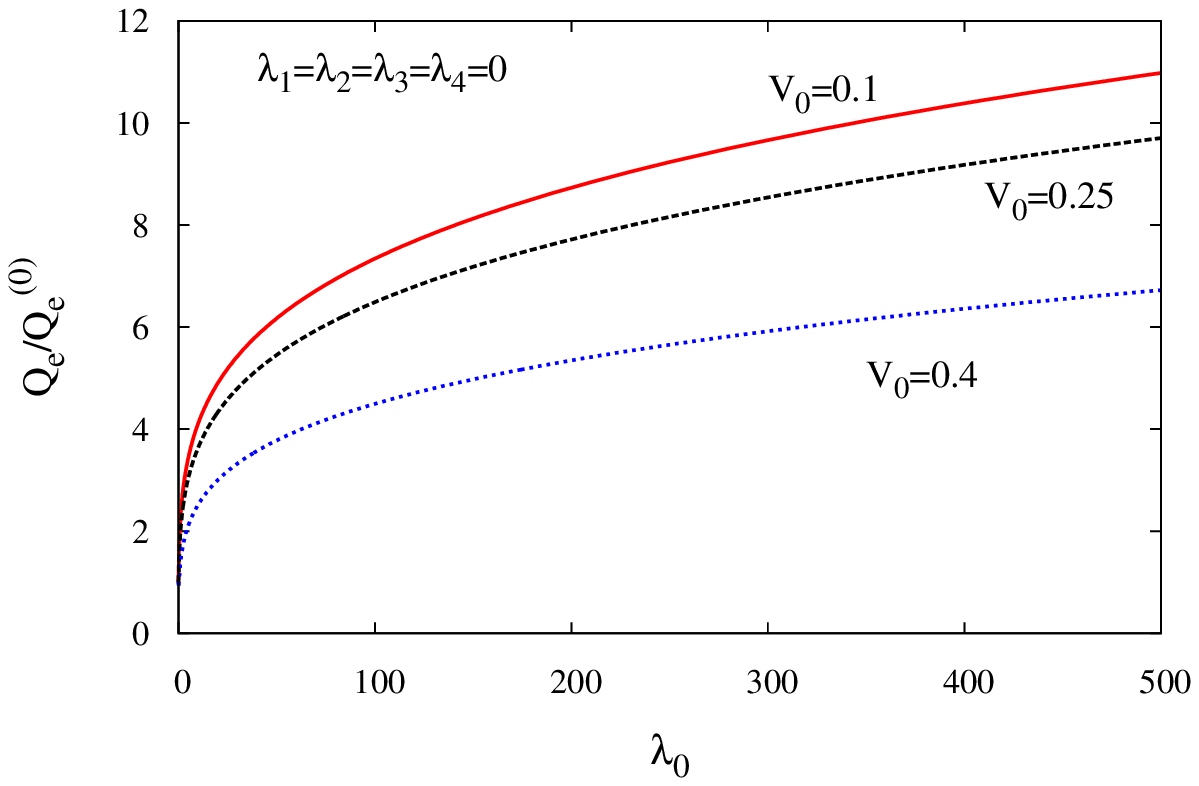,width=8cm}}
 
\end{picture}
\\

\begin{picture}(8,6)
\put(-0.5,0.0){\epsfig{file=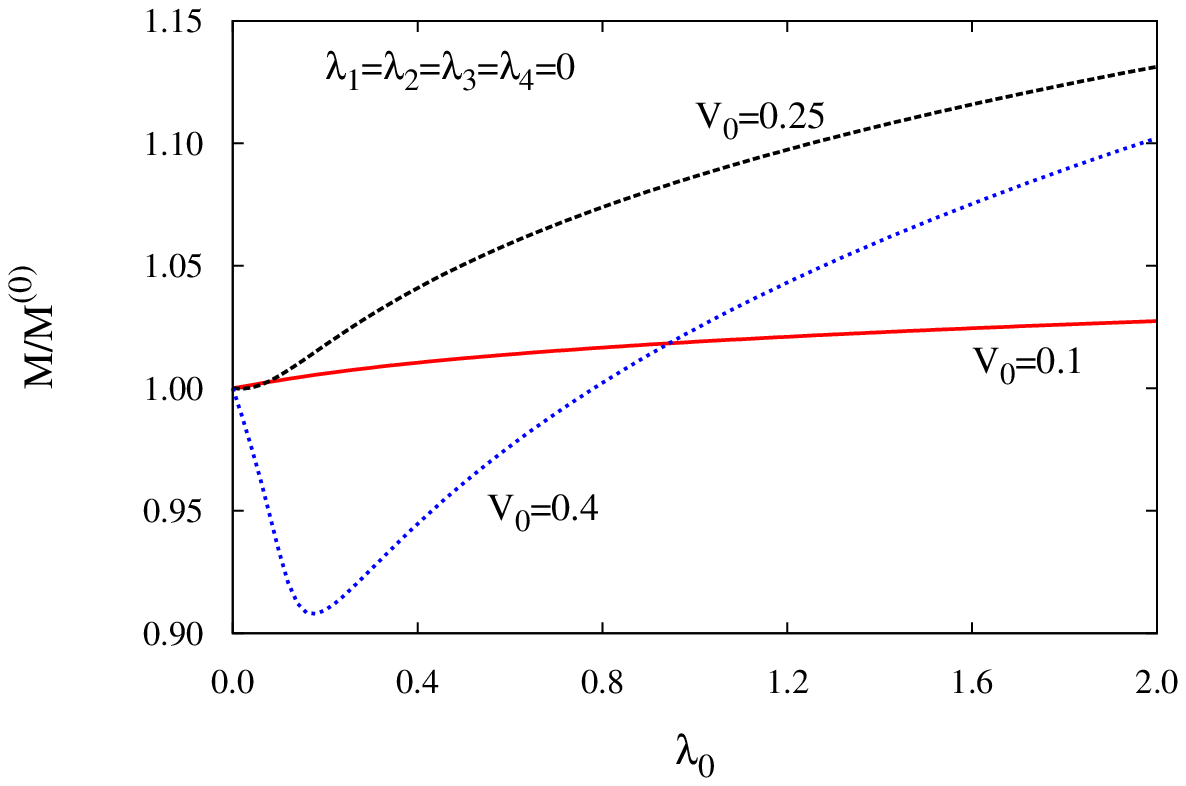,width=8cm}}
\put(8,0.0){\epsfig{file=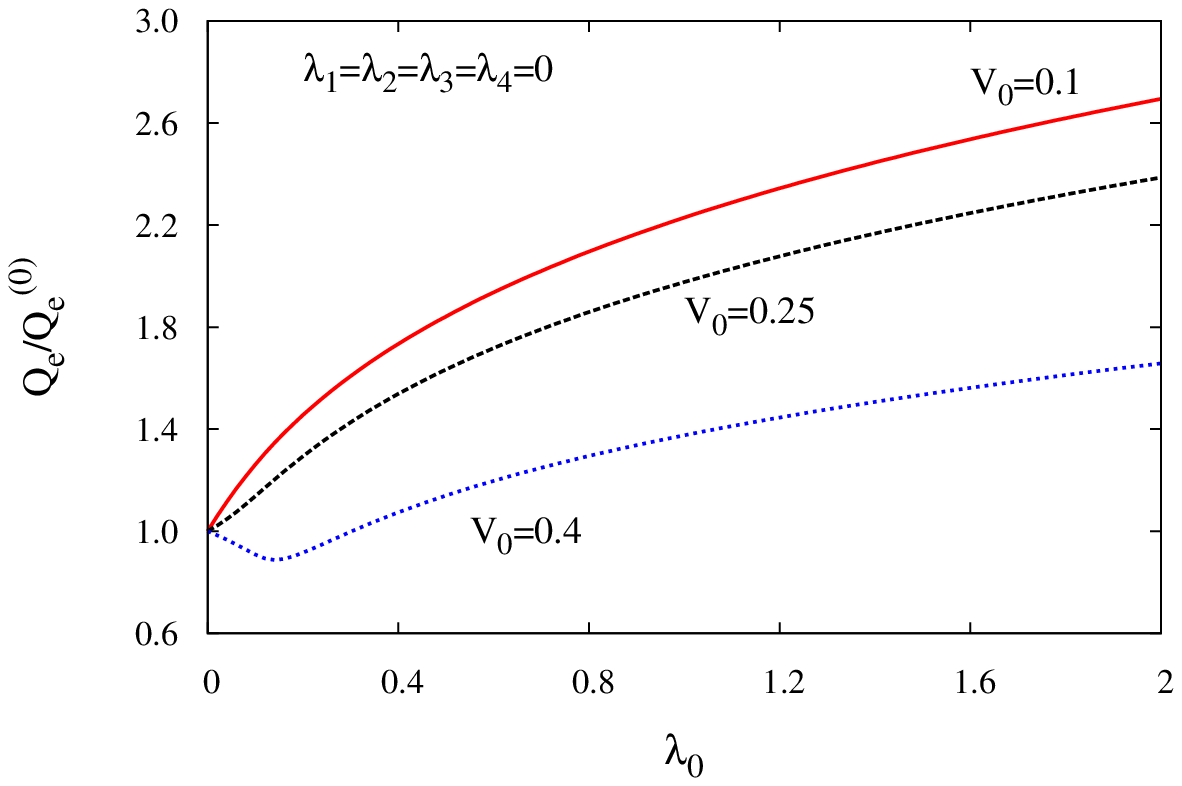,width=8cm}}
\end{picture}
  \\
{\small {\bf Figure 1.}
{\small  
The mass $M$ and the  electric charge $Q_e$ are shown
as a function of the coupling constant $\lambda_0$
for type (I) spherically symmetric generalized dyon solutions.
Several values of the electric potential $V_0$ are considered.
The bottom panel is a zoom in
plot of the top one.
Here and in Figures 2-6,  $M$ and $Q_e$
are normalized with respect to the corresponding solutions in the Georgi-Glashow model with the same $V_0$.
 }
}
\\

\setlength{\unitlength}{1cm}
\begin{picture}(8,6)
\put(-0.5,0.0){\epsfig{file= 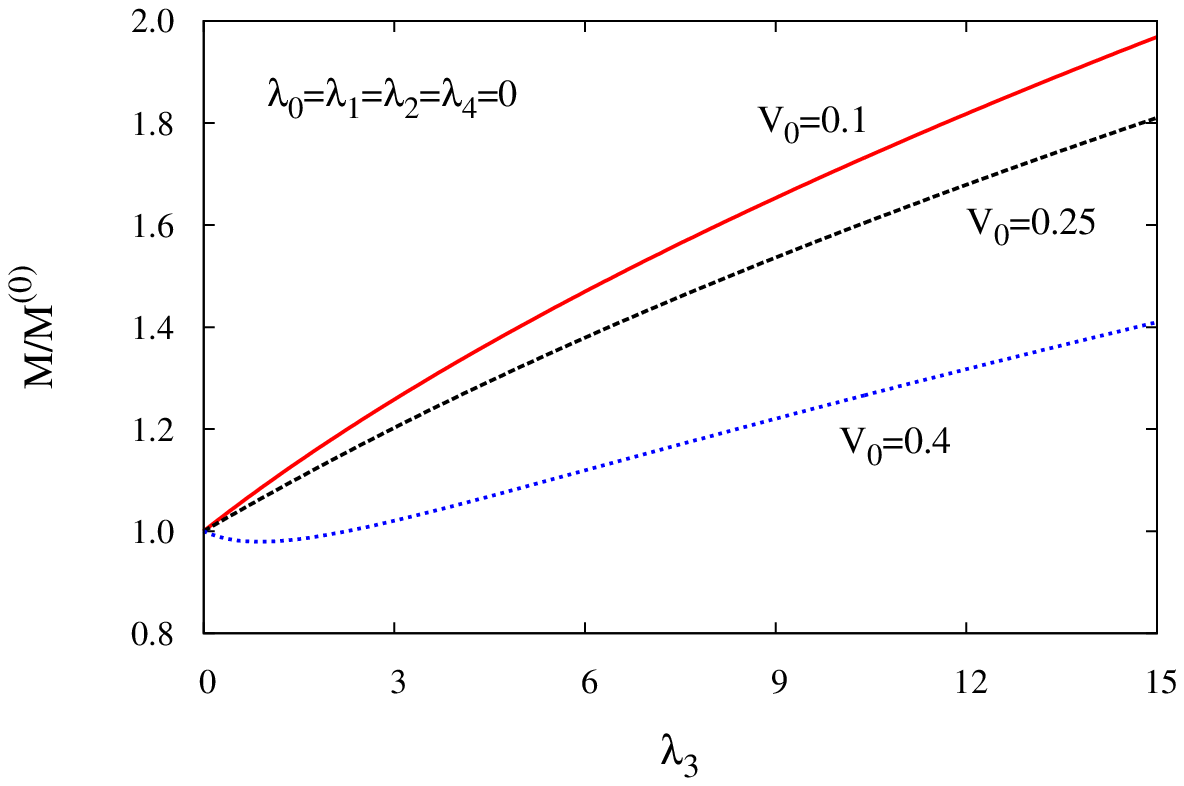,width=8cm}}
\put(8,0.0){\epsfig{file= 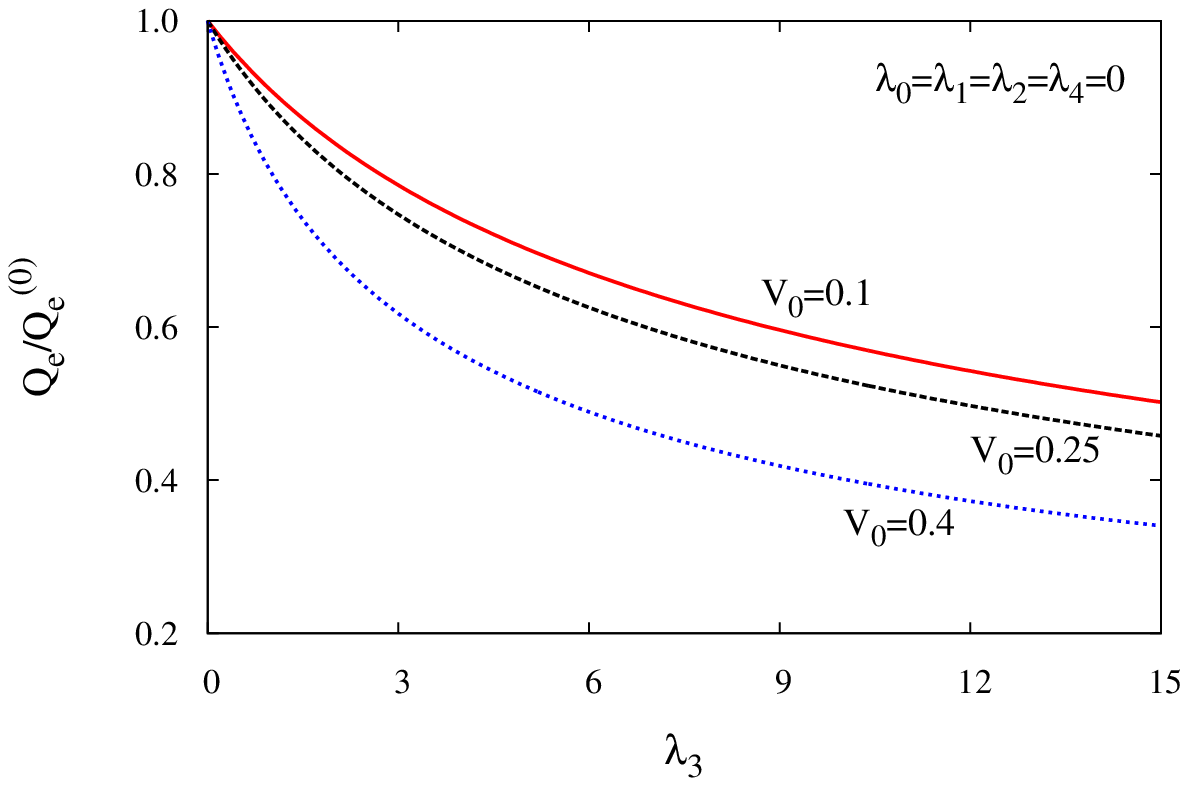,width=8cm}}
\end{picture}
  \\
{\small {\bf Figure 2.}
{\small  
The mass $M$ and the  electric charge $Q_e$ are shown
as a function of the coupling constant $\lambda_3$
for type (II) spherically symmetric generalized dyon solutions for several
values of the electric potential $V_0$.
}
}

 \newpage
\setlength{\unitlength}{1cm}
\begin{picture}(8,6)
\put(-0.5,0.0){\epsfig{file=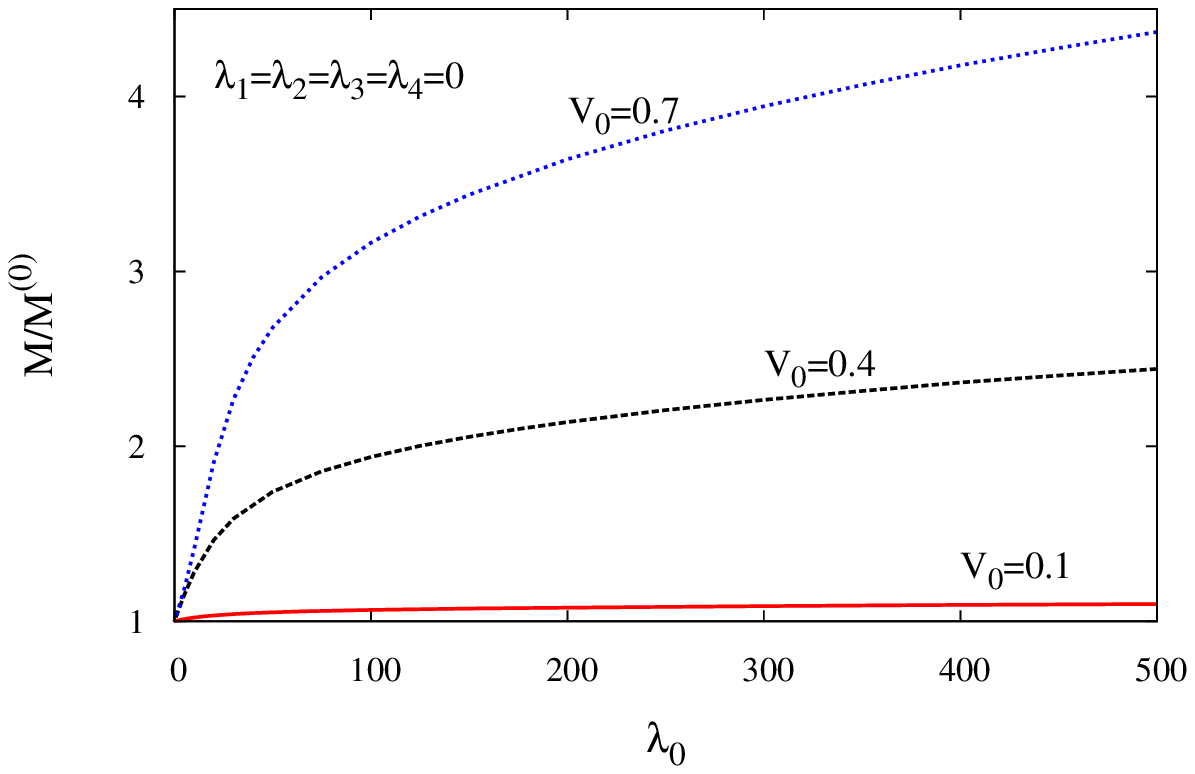,width=8cm}}
\put(8,0.0){\epsfig{file=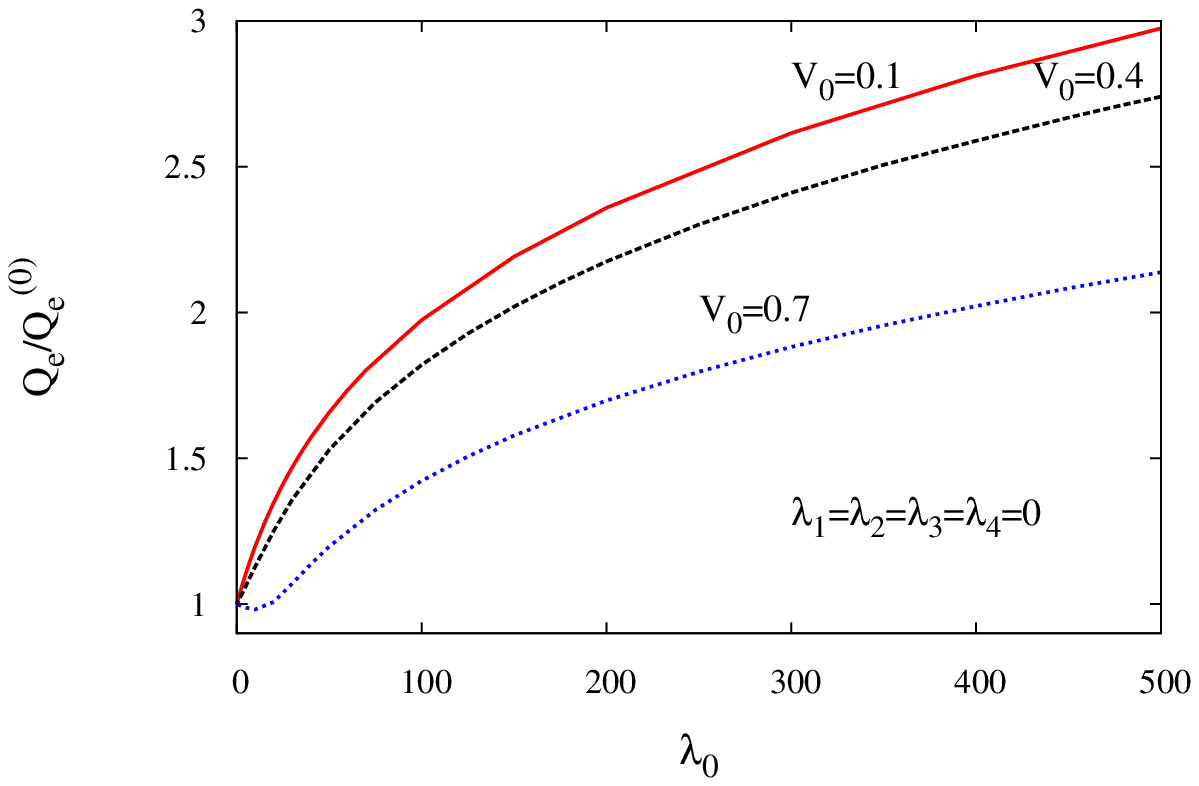,width=8cm}}
\end{picture}
  \\
{\small {\bf Figure 3.}
{\small  
The mass $M$ and the  electric charge $Q_e$ are shown
as a function of the coupling constant $\lambda_0$
for type (I) axially symmetric  generalized dyon solutions.
Several values of the electric potential $V_0$ are considered. 
The solutions here and in Figure 4 have a winding number $n=2$.
 }
} 
\vspace{1.cm}

We have studied in a systematic way the solutions 
 of the $(p=1)+(p=2)$ YMH
 model by varying the parameters $\lambda_i$ which enter the $p=2$
 reduced Lagrangian (\ref{redsphlag21}).
Here, however, we exhibit in Figures 1, 2
the results of the numerical integration
for two sub-cases of main interest only.
In both cases, we take the GG Lagrangian plus some terms in the $p=2$ Lagrangian.
Namely, we consider what we refer to as a {\bf type (I) model} 
with $\lambda_0\neq 0,\lambda_a=0$ $(a=1,\dots,4)$, and otherwise as 
a {\bf type (II) model}, 
with $\lambda_3\neq 0,\lambda_0=\lambda_1=\lambda_2=\lambda_4=0$.
The first model captures the effects of $F^4$ corrections originating in the gauge fields,
while the type (II) model involves corrections from gauged scalar fields only.
Note that the constants  $\lambda_0$ and $\lambda_3$
are made dimensionless by applying a suitable scaling.

In both cases, the properties of the 
solutions are similar to those of
their dyonic axially symmetric generalizations, and
thus we shall not discuss them further here.

\subsubsection{Axially symmetric solutions}

We turn now to the case of configurations with a nontrivial dependence of
both $r$ and $\theta$.
In our approach, the solutions to the field equations are found by
solving  numerically a set of
eight non-linear coupled partial differential equations for the functions $H_i$, $\Phi_i$
with the boundary conditions given above.
The numerical calculations
were performed with help of the package FIDISOL-CADSOL
\cite{FIDI}, based on the Newton-Raphson iterative procedure,
in which case
the known solutions of the GG model provide the initial guess.

However, the $(p=1)+(p=2)$ YMH equations with all $\lambda_i\neq 0$
are truly formidable (with some equations containing up to 500 terms),
so we eschew the general case.
Instead, we restrict our study to two sub-cases of main interest
mentioned above for the spherically symmetric limit of the model. 

The axially symmetric generalized dyons reported in this work have a winding number $n=2$.
However, we have constructed several solutions with $n=3,4$;
thus they are expected to exist for any $n$.
The considered generalized dipoles have a winding number $n=1$.
In practice, we have chosen to fix the value of $V_0$
and to compute the mass and electric charge of families of solutions
by varying   $\lambda_0$ and $\lambda_3$, respectively.
Some numerical output is  shown in Figures 3-6. 

Our results can be summarized as follows:
\begin{itemize}
\item
First, all known dyon and dipole solutions 
possess generalizations with $p=2$ terms.
However, with $\lambda_i\neq 0$, the generalized
dyon solutions satisfy the second-order
Euler--Lagrange equations and not the first order Bogomol'nyi equations as in the pure $p=1$
case.
As a result, the solutions always possess a nonvanishing angular momentum density, 
$T_\varphi^t \neq  0$ (despite the fact the total angular momentum of the 
generalized dyons vanishes).

 \vspace{1.cm} 
 \setlength{\unitlength}{1cm}
\begin{picture}(8,6)
\put(-0.5,0.0){\epsfig{file=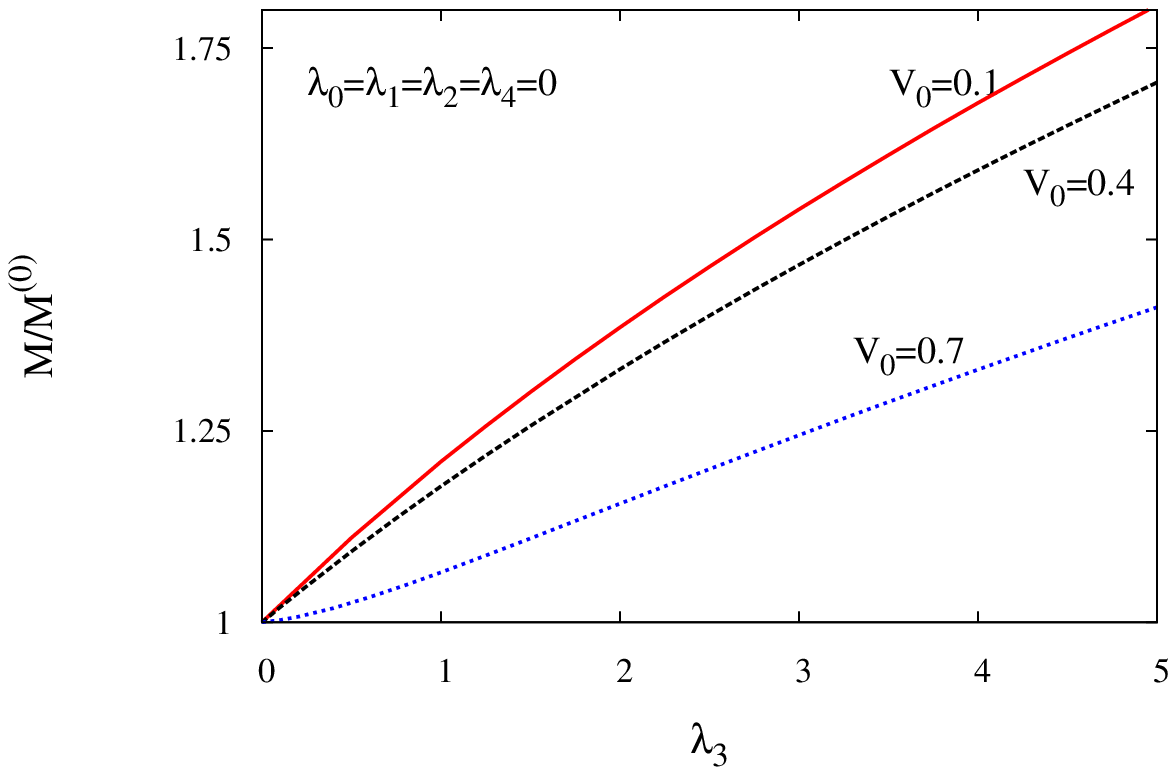,width=8cm}}
\put(8,0.0){\epsfig{file=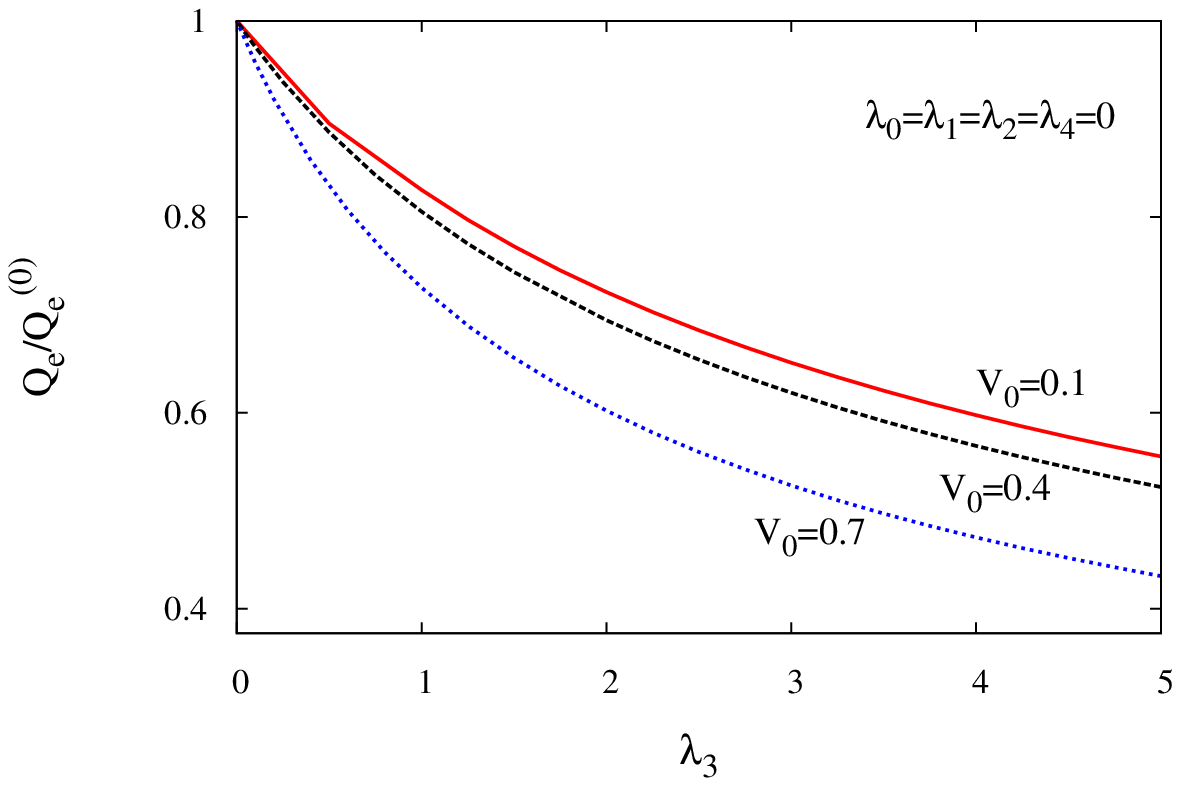,width=8cm}}
\end{picture}
 \\
{\small {\bf Figure 4.}
{\small  
The mass $M$ and the  electric charge $Q_e$
  are shown
as a function of the coupling constant $\lambda_3$
for type (II) axially symmetric  generalized dyon solutions
 and three values of
 the electric potential $V_0$.
  }
}
\\
\\
\setlength{\unitlength}{1cm}
\begin{picture}(8,6)
\put(-0.5,0.0){\epsfig{file=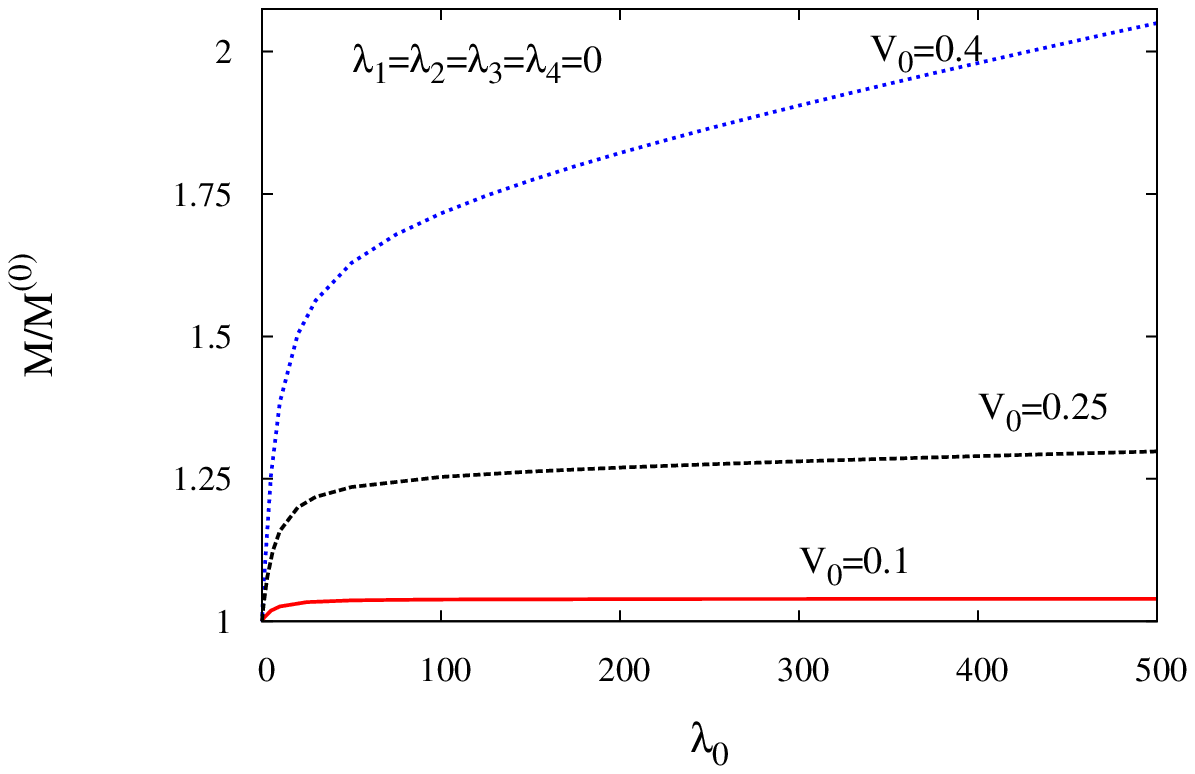,width=8cm}}
\put(8,0.0){\epsfig{file=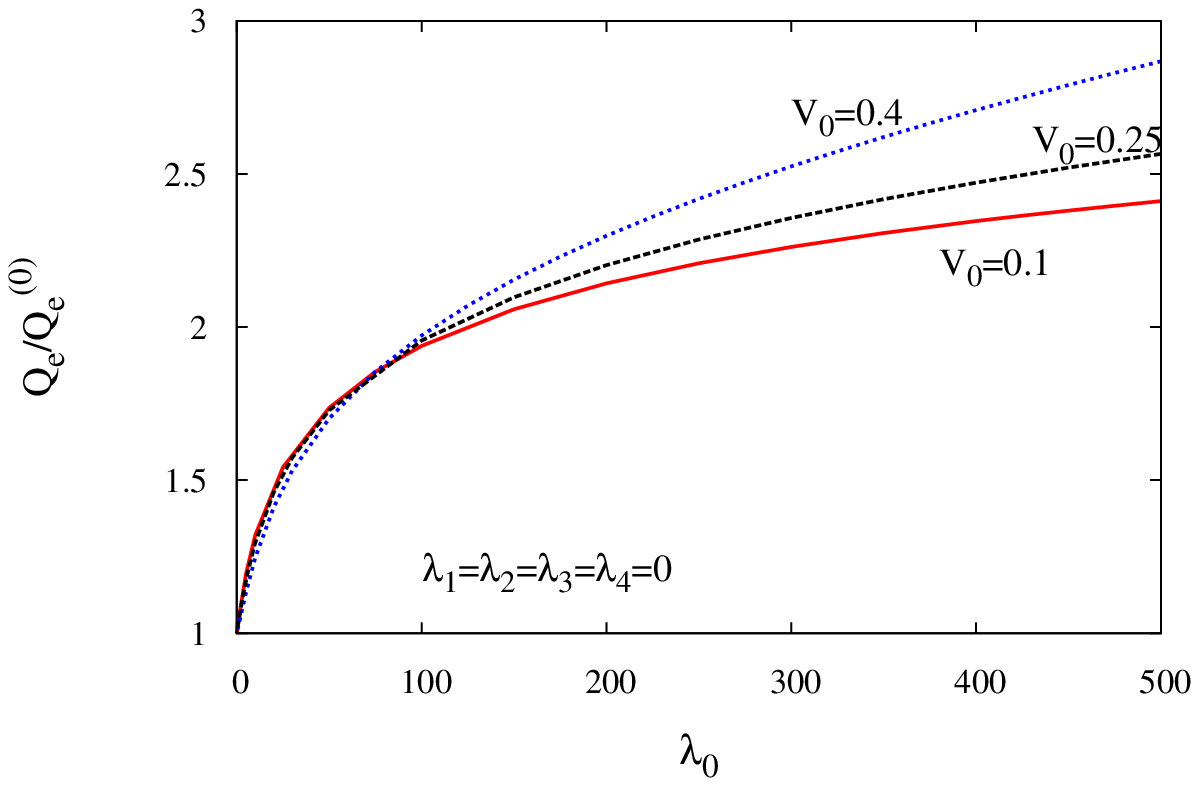,width=8cm}}
\end{picture}
  \\
{\small {\bf Figure 5.}
{\small  
Same as Figure 3 for  electrically charged, generalized dipole solutions.
Here and in Figure 6 the winding number is $n=1$.
 }
}
 \vspace{1.cm}

\item
For all solutions, the profiles of the functions $H_i$ and $\Phi_i$
look similar to the corresponding ones in the GG limit.
The same holds for the distribution of the energy and angular momentum
densities.
For generalized dyons, the energy density has a strong peak along the $\rho$ axis,
and it decreases monotonically along the symmetry axis.
Equal density contours reveal a torus-like shape of the configurations, the torii being localized in the equatorial plane. 
For the $n=1$ generalized dipoles, 
the  energy density  always possesses maxima on the positive and negative
$z$-axis at the locations of the monopole and antimonopole and a saddle
point at the origin.
Equal density contours consist in two tori on the symmetry axis. 

\item
Another  property of GG model that persists in the presence of $p=2$ corrections, is that found by
Houston and O'Raifeartaigh \cite{Houston:1980aj}: any regular axially symmetric magnetic charge distribution can be
located only at isolated points situated on the axis of symmetry, with equal and opposite values of the
charge at alternate points. In particular, if only one sign of the magnetic  charge is allowed 
($i.e.$ for generalized dyons), all the magentic  charge is
concentrated at the origin, where the Higgs field vanishes. 

\newpage
\vspace{1.cm} 
 \setlength{\unitlength}{1cm}
\begin{picture}(8,6)
\put(-0.5,0.0){\epsfig{file=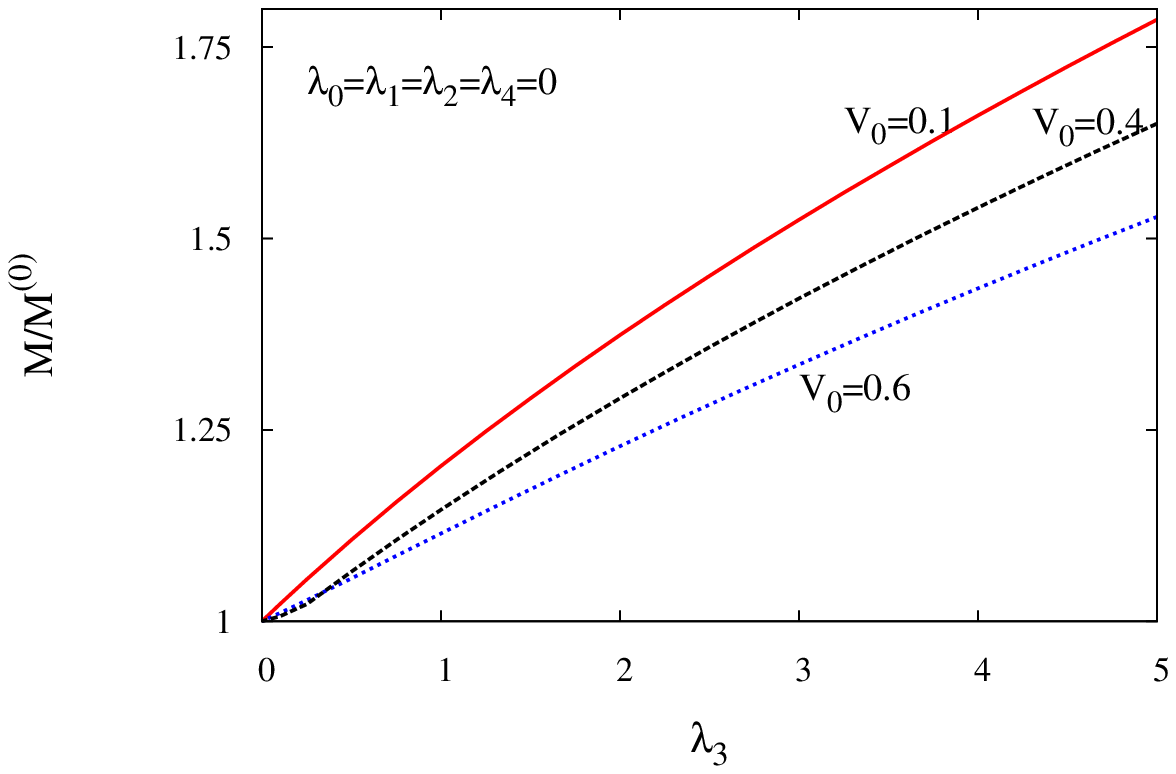,width=8cm}}
\put(8,0.0){\epsfig{file=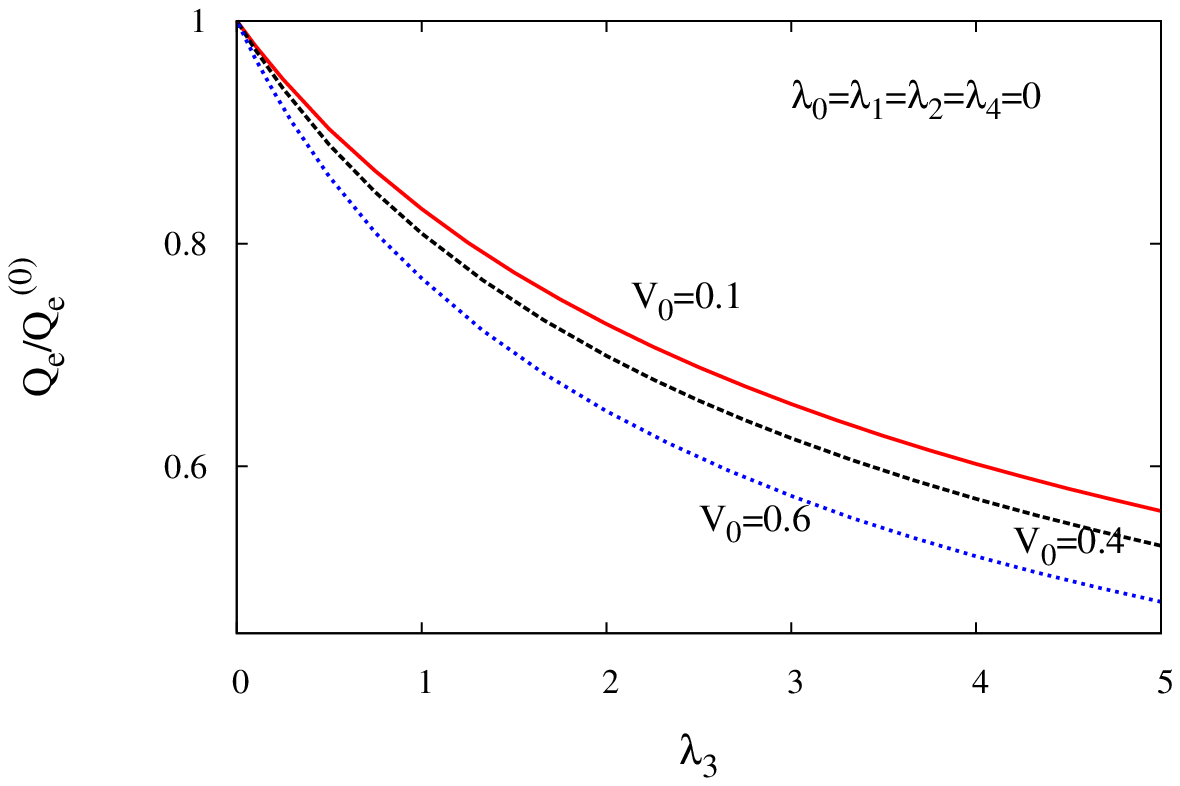,width=8cm}}
\end{picture}
 \\
{\small {\bf Figure 6.}
{\small  
Same as Figure 4 for  electrically charged, generalized dipole solutions.
 }
}
\vspace{1.cm}

\item
No upper bounds seem to exist on the values of 
the coupling constants $\lambda_0$, $\lambda_3$.
As seen in Figures 1 and 3, adding an $F^4$ term to the GG model (a type (I) model)
and taking large enough values of $\lambda_0$,
increases the mass and the electric charge of the solutions
(this feature holds for any value of 
the electric potential $V_0$).
However, rather unexpectedly, the behaviour is different for small 
$\lambda_0$ and large $V_0$,
with the existence of a minimum for both $M$ and $Q_e$  
below the values found in the GG model 
(see the Figure 1).
At the same time, while the mass increases with $\lambda_3$ for type (II) model,
the electric charge decreases 
(this holds for both generalized dyons and generalized dipoles, see Figures 4 and 6).

\end{itemize}

 More complex configurations representing 
 chains of $m$ monopoles and antimonopoles  
are known to exist for the case of a GG model \cite{Kleihaus:2003nj}.
We expect these solutions to possess also generalizations within the $(p=1)+(p=2)$ YMH model.


\section{Further remarks.} 
The main purpose of this work was to address the question on how general
the relation between angular momentum, and
electric and magnetic charges is, originally derived in \cite{VanderBij:2001nm} 
for the GG model.  
We have considered various generalizations
of the GG model, and have found that
the conjecture in~\cite{VanderBij:2001nm} 
that {\it 'a nonvanishing total angular momentum is incompatible with a net magnetic charge'}
remains valid. Perhaps most prominently, we have considered
the correction of the GG model by adding a {\it fourth order} Yang-Mills-Higgs (YMH) density, 
which also does
not result in an angular momentum for the corresponding axially symmetric generalized dyon. 
There is
no question that this qualitative conclusion will remain valid when $2p^{th}$ order YMH terms are introduced.
In addition to these, we considered the Born-Infeld--Higgs system, as well as a model \cite{Gibbons:1993xt} where
a peculiar parity-charge
violating term is added to the GG Lagrangian. In both cases precisely the same conclusion was arrived at. 
This result is
independent of the dynamics, at least for the models considered here, which we believe is exhaustive.
 It relies on the 
asymptotic behaviours of the gauge potential and Higgs field at infinity.
The mechanism behind this result turns out to be that
the presence in the Lagrangian of the usual 'quadratic kinetic' term 
$D_\mu\Phi D^\mu \Phi$ of the Higgs field enforces the same asymptotic behaviour at infinity
as in that of the GG model.

 Whether there is a possibility of circumventing this obstacle seems unlikely. Indeed, it is possible to
exclude the usual Higgs 'quadratic kinetic' term from the Lagrangian by choosing $e.g.$ 
to work exclusively with an (unphysical) model
consisting of the usual $F^2$ YM term plus the $p=2$ Lagrangian \re{LYMH2}, $i.e.$ in the absence of the usual 
$D_\mu\Phi D^\mu \Phi$ 
Higgs kinetic term.  
We have constructed such solutions, but only in the spherically symmetric limit.
In that case one finds solutions with essentially different (non-standard) asymptotics
of the magnetic gauge potential. Notably in this case the asymptotic
YM connection  $w(r)$ decays in a manner that it does not describe a magnetic monopole, but rather some other uni-pole like $e.g.$
the Skyrme hedgehog. From this, one might infer that the angular
momentum surface integral (for the corresponding axially symmetric solutions when they are found) might conceivably not vanish. 
This possibility is very unlikely because a nonvanishing angular momentum is known to be concurrent with a nonvanishing electric field,
and, when this (non-standard) spherically symmetric soliton is charged with an electric field in the manner
of Julia and Zee, the asymptotics revert to the standard asymptotics of the ``magnetic monopole'' type,
that precludes non-vanishing global angular momentum. 
We will elaborate on the details elsewhere, and have given here a
brief description in the Appendix, to support this claim. 

Concerning various possible generalizations of the results in this work,
let us mention first that the inclusion of the gravity effects does not change the 
general relations in \cite{VanderBij:2001nm}  
between the angular momentum and magnetic and electric charges,
as long as the configurations do not possess an event horizon\footnote{Moreover,
the same results hold for Anti-de Sitter asymptotics of the spacetime.
The case of a positive cosmological constant has not been
yet considered in the literature.}.
The situation changes for black hole solutions;
for example, the total angular momentum of a  
dyonic black hole solution is nozero, 
due to the contribution of the angular momentum
\cite{Kleihaus:2007vf}.
However, a discussion of these aspects is beyond the scope of the present work.

An interesting version of the problem considered in this work
is the pure YM limit, $i.e.$ no Higgs field.
In the pure $F^2$-YM theory, 
a number of well-known results forbid the existence of particle-like
soliton solutions.
In order for such configurations to exist, one has to couple the model to gravity \cite{Bartnik:1988am}
(or,  simpler, to a dilaton field \cite{Lavrelashvili:1992cp}).
Interestingly, as argued in \cite{VanderBij:2001nm},
in contrast to other solitons,
these solutions do not possess spinning generalizations.
However, considering a more general Lagrangian of the YM
 field (for example with an extra $F^4$ term
or working directly in a  Born-Infeld theory) may give 
a hope to circumvent this result (note that a non-Abelian
 Born-Infeld theory
allows for YM solitons
 without coupling to any other field \cite{Gal'tsov:1999vn}). 
 Unfortunately, this is not the case, the mechanism 
 put forward in \cite{VanderBij:2001nm} for a $F^2$ theory
 being still valid in this case.
Again, the problem resides in the asymptotic behaviour of the gauge potentials:
since the $F^2$ contribution would dominate
in the field equations as $r\to \infty$,
the electric components of the
gauge field act like an isotriplet Higgs field with negative metric, and by themselves cause some of the magnetic
components to oscillate rather than decrease exponentially, which would lead to delocalized configurations.
Therefore, we are forced to take a vanishing asymptotic value of the electric gauge potential,  
$V_0=0$. 
However, from (\ref{1}) this implies $A_t\equiv 0$ (since ${\cal E}\geq 0$), $i.e.$
a static configuration.

Finally, let us address the question of spinning solutions in the standard model. The results in
\cite{Radu:2008ta},
\cite{Kleihaus:2008cv}
show that the well-known Klinkhamer-Manton
sphalerons possess spinning generalizations.
Moreover, for a nonzero mixing angle $\theta_W$,
the angular momentum of a spinning sphaleron equals the electric charge
\cite{Radu:2008ta},
\cite{Kleihaus:2008cv} 
(note the analogy with the monopole-antimonopole pair).
However, the Lagrangian of the Weinberg-Salam model
can be supplemented with higher derivative corrections,
the simplest one consisting in Skyrme-like terms
of the Higgs field  plus  $F^4$ terms of the gauge fields.
Based on the results in this work, we conjecture that
the qualitative results in
\cite{Radu:2008ta},
\cite{Kleihaus:2008cv}
remain always valid and the angular momentum of a spinning sphaleron
always equals the electric charge. 
 
 \vspace*{0.7cm}
\noindent{\textbf{~~~Acknowledgements.--~}} 
D.H. Tch. is grateful to the Albert-Einstein-Institut (Golm) for their
hospitality, where this work was started. 
E.R. gratefully acknowledges support from the FCT-IF programme. F.N-L. acknowledges support from the Spanish Education and Science Ministry under Project No. FIS2011-28013.

\appendix
 
\section*{Appendix A: Solitons of the $F^2$ plus $p=2$ YMH system in $3+1$ dimensions}
\setcounter{equation}{0} 
\renewcommand{\theequation}{A.\arabic{equation}}
 
In this Appendix we consider a model consisting of the usual $F^2$ Yang-Mills
term plus the pure $p=2$ YMH system in $3+1$ dimensions. 
Our original aim was to find solutions with a more general asymptotic
behaviour of the gauge potentials,
in the hope of circumventing the ban on spinning YMH monopole solitons.
As this attempt gave a final negative result, the analysis was
relegated to the Appendix, in support of our claim that topologically stable YMH monopoles cannot spin.

It is important in this context to distinguish between a topologically stable monopole (or uni-pole) and a topologically stable ``magnetic monopole''.
The hedgehog of the Skyrme model is a monopole centred at the origin. The 't~Hooft-Polyakov hedgehog on the other hand is a monopole,
with the additional property that its asymptotic gauge field is a Dirac-Yang $U(1)$ Maxwell field. 
In this case
the magnetic gauge function $w(r)$ 
vanishes asymptotically,
$i.e.$ $w(r)=0$ as $r\to\infty$. 
By contrast, the solutions to the system
\be
\label{YM+p=2}
{\cal L}=-\frac{1}{4}\alpha \mbox{Tr} \left( F_{\mu\nu}\,F^{\mu\nu} \right)+{\cal L}^{(2)},
\ee
with ${\cal L}^{(2)}$ given by \re{LYMH2} 
have $\lim_{r\to\infty}w(r)\neq 0$.


Before discussing this  case, let us consider first 
the pure $p=2$ YMH magnetic system ($i.e.$ without a $F^2$ term, $\alpha=0$).
The corresponding energy density functional is found by taking in (\ref{redsphlag21}) 
$u(r)=0$
together with
 $\lambda_0=0$, $\lambda_1=\lambda_2=\lambda_3= \lambda_4=1$,
 and reads (up to some unimportant overall numerical factor):
\bea
H^{(2)}_{\rm mag}&=&\eta^2 \left([(1-w^2)h]'\right)^2+3^2\,\eta^8\,r^2\,(1-h^2)^4\nonumber\\
&&+2^2\eta^4\left([(1-h^2)w]'\right)^2+3^2\cdot 2^4\,\eta^6\,(1-h^2)^2\,w^2h^2\label{energyp=2YMH}\\
&&+3^2\cdot 2^3\,\eta^6\,r^2\,(1-h^2)^2\,h'^2+2\,\eta^4\,r^{-2}[(1-h^2)(1-w^2)+2\,w^2h^2]^2,
\nonumber
\eea
being bounded from below by the topological charge density \cite{Kleihaus:1998kd}
\be
\label{lb1}
\rho=2\,\eta^5\,\frac{d}{dr}\{(h-\frac23h^3+\frac15h^5)-(1-h^2)^2\,w^2h\}.
\ee
However, this bound is never achieved, since the 
 Bogomol'nyi equations of the pure $p=2$ model
\bea
\nonumber
\eta\,r^{-1}[(1-w^2)h]'&=&\pm3\,\eta^4\,r\,(1-h^2)^2,
\label{bog1}
\\
\eta^2\,[(1-h^2)w]'&=&\mp3!\,\eta^3\,(1-h^2)\,wh,
\label{bog2}
\\
3!\,\eta^3\,r\,(1-h^2)\,h'&=&\pm\eta^2\,r^{-1}[(1-h^2)(1-w^2)+2\,w^2h^2],
\label{bog3}
\nonumber
\eea
are overdetermined \cite{Tchrakian:1990gc}.

Returning to the case of the model
\[
H=\frac14\,\al\,r^2\,\left[2\left(\frac{w'}{r}\right)^2+\frac{1}{r^4}(1-w^2)^2\right] +H^{(2)}_{\rm mag},
\]
 with arbitrary $\lambda_i$ (and $u(r)=0$),
one notices that the regularity of the solutions imposes
the same boundary conditions at $r=0$ as in the $p=1$ case, $i.e.$ $w(0)=1$ and $h(0)= 0$.
However, the far field asymptotics can be different,
the magnetic gauge potential $w(r)$ 
 possesing another asymptotic value compatible with finite energy requirement, apart from the standard one $w(\infty)=0$.
We have found numerical evidence for the existence
of solutions smoothly interpolating between $w(0)=1$ and $h(0)= 0$
and
\begin{eqnarray}
\label{new-as}
 h(r) \to 1,~~ w(r)\to \frac{1}{\sqrt{1+192  {\eta^4\lambda_2}/{\alpha} }}~~{\rm as}~~r\to \infty.
\end{eqnarray} 
Their total mass is finite, possesing a nontrival dependence on 
the parameters $\lambda_i$.

The situation is, however, different when looking for electric generalizations  
of these configurations.
It turns out that no solutions compatible with the finite energy requirements
can be found for the asymptotics (\ref{new-as}).
This can be understood as follows.
First, the energy density possess a $u^2 w^2$ term, which
originates in the $F^2$-part of the Lagrangian.
This term should decay faster than $1/r$ as $r\to \infty$;
thus a value $w(\infty)\neq 0$
results in $u(\infty)=V_0= 0$.
However, this implies directly a vanishing
electric potential, as seen from the relation (\ref{E1}),
which still holds in this case.
Thus we conclude that a nonzero electric potential 
is not compatible with non-standard asymptotics
of the magnetic gauge potential in the $F^2$ plus $p=2$ YMH system. 

On general grounds,
we expect a similar result to hold as well
in the axially symmetric case.

\newpage
\begin{small}

\end{small}


\begin{thebibliography}{99}
\bibitem{book}
N. S. Manton and P. Sutcliffe, {\it Topological solitons}, Cambridge, UK: Univ. Pr. (2004).
\bibitem{Skyrme}
  T.~H.~R.~Skyrme,
  Proc.\ Roy.\ Soc.\ Lond.\ A {\bf 260} (1961) 127.
  \\
  T.~H.~R.~Skyrme,
  Nucl.\ Phys.\  {\bf 31} (1962) 556.
\bibitem{Deser:1976wq}
  S.~Deser,
  Phys.\ Lett.\ B {\bf 64} (1976) 463.
\bibitem{Bartnik:1988am}
  R.~Bartnik and J.~Mckinnon,
  Phys.\ Rev.\ Lett.\  {\bf 61} (1988) 141.
  
\bibitem{'tHooft:1974qc}
  G.~'t Hooft,
  Nucl.\ Phys.\ B {\bf 79} (1974) 276.
\bibitem{Polyakov:1974ek}
  A.~M.~Polyakov,
  JETP Lett.\  {\bf 20} (1974) 194
   [Pisma Zh.\ Eksp.\ Teor.\ Fiz.\  {\bf 20} (1974) 430].
\bibitem{Tchrakian:2010ar}
  T.~Tchrakian,
  J.\ Phys.\ A {\bf 44} (2011) 343001
  [arXiv:1009.3790 [hep-th]].
\bibitem{Tch} 
D.H. Tchrakian, J. Math. Phys, {\bf 21} (1980) 166;
\\
D.H. Tchrakian, 
 Phys. Lett. B {\bf 150} (1985) 360;
 \\
 D.H. Tchrakian, 
  Int. J. Mod. Phys. A (Proc. Suppl.) {\bf 3A} (1993) 584.  
\bibitem{Klinkhamer:1984di}
  F.~R.~Klinkhamer and N.~S.~Manton,
  Phys.\ Rev.\ D {\bf 30} (1984) 2212.
\bibitem{JZ}
  B.~Julia and A.~Zee,
  Phys.\ Rev.\ D {\bf 11} (1975) 2227.
    
\bibitem{VanderBij:2001nm}
  J.~J.~Van der Bij and E.~Radu,
  Int.\ J.\ Mod.\ Phys.\ A {\bf 17} (2002) 1477
  [gr-qc/0111046];
  \\
  J.~J.~van der Bij and E.~Radu,
  Int.\ J.\ Mod.\ Phys.\ A {\bf 18} (2003) 2379
  [hep-th/0210185].
\bibitem{Kleihaus:2005fs}
  B.~Kleihaus, J.~Kunz and U.~Neemann,
  Phys.\ Lett.\ B {\bf 623} (2005) 171
  [gr-qc/0507047].
\bibitem{Paturyan:2004ps}
  V.~Paturyan, E.~Radu and D.~H.~Tchrakian,
  Phys.\ Lett.\ B {\bf 609} (2005) 360
  [hep-th/0412011].
\bibitem{AyonBeato:1998ub}
  E.~Ayon-Beato and A.~Garcia,
  Phys.\ Rev.\ Lett.\  {\bf 80} (1998) 5056
  [gr-qc/9911046].
\bibitem{Gal'tsov:1999vn}
  D.~Gal'tsov and R.~Kerner,
  Phys.\ Rev.\ Lett.\  {\bf 84} (2000) 5955
  [hep-th/9910171].
\bibitem{Radu:2011ip}
  E.~Radu and D.~H.~Tchrakian,
  Phys.\ Rev.\ D {\bf 85} (2012) 084022
  [arXiv:1111.0418 [gr-qc]].
 
  
\bibitem{Navarro-Lerida:2013pua}
  F.~Navarro-Lerida and D.~H.~Tchrakian,
  arXiv:1311.3950 [hep-th].
  
\bibitem{Volkov:2003ew}
  M.~S.~Volkov and E.~Wohnert,
  Phys.\ Rev.\ D {\bf 67} (2003) 105006
  [hep-th/0302032].
\bibitem{Heusler:1996ft}
M.~Heusler,
Helv.\ Phys.\ Acta {\bf 69} (1996) 501.
\bibitem{Forgacs:1980zs}
P.~Forgacs and N.~S.~Manton,
Commun.\ Math.\ Phys.\ {\bf 72} (1980) 15;
\\
P.~G.~Bergmann and E.~J.~Flaherty,
J.\ Math.\ Phys.\ {\bf 19} (1978) 212.
\bibitem{LL}
  L.~D.~Landau, E.~M.~Lifshitz, 
   {\it 'Course of  Theoretical Physics. Vol. 2: The Classical Theory of Fields',}
  Butterworth-Heinemann (1975), UK, 1975.
\bibitem{Belinfante} 
  F. J. Belinfante,
  Physica {\bf 6} (1939) 887;
  \\
  F. J. Belinfante,  
  Physica {\bf 7}  (1940) 449.
\bibitem{Volkov:1998cc}
  M.~S.~Volkov and D.~V.~Gal'tsov,
  Phys.\ Rept.\  {\bf 319} (1999) 1
  [hep-th/9810070].
\bibitem{Maleknejad:2011jw}
  A.~Maleknejad and M.~M.~Sheikh-Jabbari,
  Phys.\ Lett.\ B {\bf 723} (2013) 224
  [arXiv:1102.1513 [hep-ph]].
\bibitem{Tseytlin:1997csa}
  A.~A.~Tseytlin,
  Nucl.\ Phys.\ B {\bf 501} (1997) 41
  [hep-th/9701125];
 \bibitem{Gauntlett:1997ss}
  J.~P.~Gauntlett, J.~Gomis and P.~K.~Townsend,
  JHEP {\bf 9801} (1998) 003
  [hep-th/9711205];
\\
  D.~Brecher and M.~J.~Perry,
  Nucl.\ Phys.\ B {\bf 527} (1998) 121
  [hep-th/9801127].
\bibitem{Grandi:1999rv}
  N.~E.~Grandi, E.~F.~Moreno and F.~A.~Schaposnik,
  Phys.\ Rev.\ D {\bf 59} (1999) 125014
  [hep-th/9901073].
\bibitem{Grandi:1999dv}
  N.~E.~Grandi, A.~Pakman, F.~A.~Schaposnik and G.~A.~Silva,
  Phys.\ Rev.\ D {\bf 60} (1999) 125002
  [hep-th/9906244].

\bibitem{Radu:2011zy}
  E.~Radu and T.~Tchrakian,
{\it 'New Chern-Simons densities in both odd and even dimensions'}, in
{\it Low Dimensional Physics and Gauge Principles. Matinyan's Festschrift}, eds. V.G. Gurzadyan, A. Kl\"umper, A.G. Sedrakyan,
World Scientific, 2013
  [arXiv:1101.5068 [hep-th]].
\bibitem{Gibbons:1993xt}
  G.~W.~Gibbons, D.~Kastor, L.~A.~J.~London, P.~K.~Townsend and J.~H.~Traschen,
  Nucl.\ Phys.\ B {\bf 416} (1994) 850
  [hep-th/9310118].
\bibitem{7}
N.S. Manton, Nucl.Phys.{\bf B135} (1978) 319. 
\bibitem{Rebbi:1980yi}
  C.~Rebbi and P.~Rossi,
  Phys.\ Rev.\ D {\bf 22} (1980) 2010.
\bibitem{Brihaye:1994ib}
  Y.~Brihaye and J.~Kunz,
  Phys.\ Rev.\ D {\bf 50} (1994) 4175
  [hep-ph/9403392].
\bibitem{Radu:2004ys}
  E.~Radu and D.~H.~Tchrakian,
  Phys.\ Rev.\ D {\bf 71} (2005) 064002
  [hep-th/0411084].
\bibitem{Kleihaus:2003nj}
  B.~Kleihaus, J.~Kunz and Y.~Shnir,
  Phys.\ Lett.\ B {\bf 570} (2003) 237
  [hep-th/0307110].
\bibitem{Kleihaus:1998kd}
  B.~Kleihaus, D.~O'Keeffe and D.~H.~Tchrakian,
  Phys.\ Lett.\ B {\bf 427} (1998) 327.
\bibitem{Kleihaus:1998gy}
  B.~Kleihaus, D.~O'Keeffe and D.~H.~Tchrakian,
  Nucl.\ Phys.\ B {\bf 536} (1998) 381
  [hep-th/9806088].
\bibitem{Tchrakian:1990gc}
  D.~H.~Tchrakian and A.~Chakrabarti,
  J.\ Math.\ Phys.\  {\bf 32} (1991) 2532.
\bibitem{Taubes:1982ie}
C.~H.~Taubes,
Commun.\ Math.\ Phys.\  {\bf 86} (1982) 257.
\bibitem{Kleihaus:1999sx}
  B.~Kleihaus and J.~Kunz,
  Phys.\ Rev.\ D {\bf 61} (2000) 025003
  [hep-th/9909037].
\bibitem{Prasad:1975kr}
  M.~K.~Prasad and C.~M.~Sommerfield,
  Phys.\ Rev.\ Lett.\  {\bf 35} (1975) 760.
\bibitem{Hartmann:2000ja}
  B.~Hartmann, B.~Kleihaus and J.~Kunz,
  Mod.\ Phys.\ Lett.\ A {\bf 15} (2000) 1003
  [hep-th/0004108].
\bibitem{COLSYS}
 U. Ascher, J. Christiansen, R.~D. Russell,
 Mathematics of Computation 33 (1979) 659;
 ACM Transactions 7 (1981) 209.    
\bibitem{FIDI}
 W.~Sch\"onauer, and R.~Wei\ss,
 J.~Comput. Appl. Math. {\bf 27} (1989) 279;\\
 M.~Schauder, R.~Wei\ss, and W.~Sch\"onauer,
{\it  The CADSOL Program Package}, Universit\"at Karlsruhe,
 Interner Bericht Nr. 46/92 (1992).

\bibitem{Kleihaus:2000sx}
B.~Kleihaus and J.~Kunz,
Phys.\ Rev.\ {\bf D 61} (2000) 025003
[hep-th/9909037].
 
\bibitem{Houston:1980aj}
  P.~Houston and L.~O'Raifeartaigh,
  Z.\ Phys.\ C {\bf 8} (1981) 175.
\bibitem{Kleihaus:2007vf}
  B.~Kleihaus, J.~Kunz, F.~Navarro-Lerida and U.~Neemann,
  Gen.\ Rel.\ Grav.\  {\bf 40} (2008) 1279
  [arXiv:0705.1511 [gr-qc]].
\bibitem{Lavrelashvili:1992cp}
  G.~V.~Lavrelashvili and D.~Maison,
  Phys.\ Lett.\ B {\bf 295} (1992) 67;
  \\
  P.~Bizon,
  Phys.\ Rev.\ D {\bf 47} (1993) 1656
  [hep-th/9209106].
  
\bibitem{Radu:2008ta}
  E.~Radu and M.~S.~Volkov,
  Phys.\ Rev.\ D {\bf 79} (2009) 065021
  [arXiv:0810.0908 [hep-th]].
\bibitem{Kleihaus:2008cv}
  B.~Kleihaus, J.~Kunz and M.~Leissner,
  Phys.\ Lett.\ B {\bf 678} (2009) 313
  [arXiv:0810.1142 [hep-ph]].
\bibitem{Paul:1986ix}
  S.~K.~Paul and A.~Khare,
  Phys.\ Lett.\ B {\bf 174} (1986) 420
   [Erratum-ibid.\  {\bf 177B} (1986) 453].
\bibitem{Hong:1990yh}
  J.~Hong, Y.~Kim and P.~Y.~Pac,
  Phys.\ Rev.\ Lett.\  {\bf 64} (1990) 2230.
\bibitem{Jackiw:1990aw}
  R.~Jackiw and E.~J.~Weinberg,
  Phys.\ Rev.\ Lett.\  {\bf 64} (1990) 2234.
  
  
  











 
  
  
  
   
  
  



  

 




  

 




  

 




  

 




  

 




  

 




  

 

  \end{thebibliography}
\end{document}